\documentclass{styles/vgtc}                          % final (conference style)
\ifpdf%                                % if we use pdflatex
  \pdfoutput=1\relax                   % create PDFs from pdfLaTeX
  \usepackage{graphicx}                % allow us to embed graphics files
  \DeclareGraphicsExtensions{.pdf,.png,.jpg,.jpeg} % for pdflatex we expect .pdf, .png, or .jpg files
\else%                                 % else we use pure latex
  \usepackage{graphicx}                % allow us to embed graphics files
  \DeclareGraphicsExtensions{.eps}     % for pure latex we expect eps files
\fi%

\graphicspath{{figures/}{pictures/}{images/}{./}} % where to search for the images

\usepackage{microtype}                 % use micro-typography (slightly more compact, better to read)
\PassOptionsToPackage{warn}{textcomp}  % to address font issues with \textrightarrow
\usepackage{textcomp}                  % use better special symbols
\usepackage{mathptmx}                  % use matching math font
\usepackage{times}                     % we use Times as the main font
\usepackage{cite}                      % needed to automatically sort the references
\usepackage{tabu}                      % only used for the table example
\usepackage{booktabs}                  % only used for the table example
\usepackage{styles/review}
\usepackage{amsmath}

\newcommand{\addIcon}[1] {
    \raisebox{-0.25em}{\includegraphics[height=1.25em]{#1}}
}
\usepackage{xspace}

\newcommand{\inlineIcon}[1]{\hspace{-0.1mm}\includegraphics[height=0.8em]{#1}\xspace}

\newcommand{\connect}{\inlineIcon{icons/connect}}
\newcommand{\connectiveFilter}{\inlineIcon{icons/connectiveFilter}}
\newcommand{\convert}{\inlineIcon{icons/convert}}
\newcommand{\deriveAttribute}{\inlineIcon{icons/deriveAttribute}}
\newcommand{\deriveConnectedAttribute}{\inlineIcon{icons/deriveConnectedAttribute}}
\newcommand{\edgeProjection}{\inlineIcon{icons/edgeProjection}}
\newcommand{\facet}{\inlineIcon{icons/facet}}
\newcommand{\filter}{\inlineIcon{icons/filter}}
\newcommand{\promote}{\inlineIcon{icons/promote}}

\newcommand{\rollup}{\inlineIcon{icons/rollup}}
\newcommand{\supernode}{\inlineIcon{icons/supernode}}
\newcommand{\toggleDirection}{\inlineIcon{icons/toggleDirection}}

\definecolor{cb_green}{rgb}{0.105,0.62,0.467}

\usepackage{relsize}

\onlineid{0}

\vgtccategory{Research}

\vgtcinsertpkg

\title{%Origraph: Flexible Authoring and Reshaping of Networks
    Origraph: Interactive Network Wrangling
  }

\author{Alex Bigelow, Carolina Nobre, Miriah Meyer, Alexander Lex}
\authorfooter{
\item Alex Bigelow, Carolina Nobre, Miriah Meyer, and Alexander Lex are with the University of Utah. E-mail: \{abigelow, miriah\}@cs.utah.edu, \{cnobre, alex\}@sci.utah.edu.
}

\shortauthortitle{Bigelow \MakeLowercase{\textit{et al.}}: Origraph: Interactive Network Wrangling}
\abstract{Networks are a natural way of thinking about many datasets. The data on which a network is based, however, is rarely collected in a form that suits the analysis process, making it necessary to create and reshape networks. Data wrangling is widely acknowledged to be a critical part of the data analysis pipeline, yet interactive network wrangling has received little attention in the visualization research community. 
In this paper, we discuss a set of operations that are important for wrangling network datasets and introduce a visual data wrangling tool, Origraph, that enables analysts to apply these operations to their datasets. Key operations include creating a network from source data such as tables, reshaping a network by introducing new node or edge classes, filtering nodes or edges, and deriving new node or edge attributes. Our tool, Origraph, enables analysts to execute these operations with little to no programming, and to immediately visualize the results. Origraph provides views to investigate the network model, a sample of the network, and node and edge attributes. In addition, we introduce interfaces designed to aid analysts in specifying arguments for sensible network wrangling operations. We demonstrate the usefulness of Origraph in two Use Cases: first, we investigate gender bias in the film industry, and then the influence of money on the political support for the war in Yemen.} % end of abstract


\CCScatlist{ % not used in journal version
 \CCScat{Human-centered computing}{Information visualization};
 \CCScat{Human-centered computing}{Visualization systems and tools};
 \CCScat{Information systems}{Graph-based database models}
}

\teaser{
  \centering
  \includegraphics[width=\linewidth]{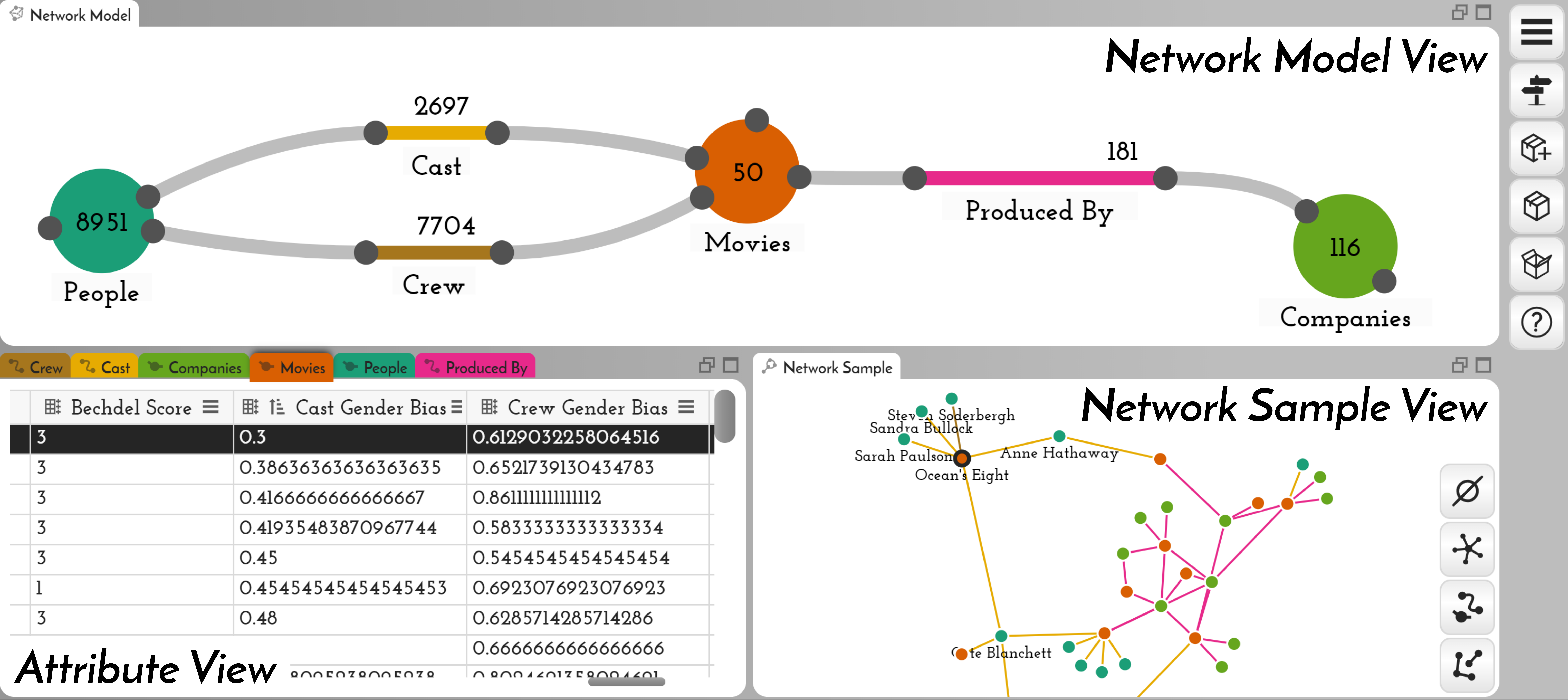}
  \vspace{-6mm}
  \caption{Overview of the Origraph UI. The \textbf{network model view} shows relationships between node and edge classes and is the primary interface for operations related to connectivity. The \textbf{attribute view} shows node and edge attributes in a table and is the primary interface for attribute-related operations. The \textbf{network sample view} visualizes a preview of the current state of the network.}
  \label{fig:teaser}
}

\begin{document}


\firstsection{Introduction}

\maketitle

%% \section{Introduction} %for journal use above \firstsection{..} instead

%\section{Terms}

%Use: Network, not Graph; 
%Use: Node; not vertex;
%Use: Edge; not link;
%Class: A class defines a common set of attributes; generic term for both nodes and edges.
%Item: a generic node/edge

%It is exceedingly rare that an analysis question can be directly answered from a readily available dataset. More often than not, analysts will use datasets as imperfect proxies that can be useful for shedding light on a question of interest~\cite{fisher_making_2018}. 

Data wrangling---which includes cleaning data, merging datasets, and transforming representations---is known to be a tedious and time-consuming part of data analysis~\cite{kandel_research_2011}. Historically, wrangling was done with scripting languages such as Python, Perl, and R, or manipulation in spreadsheet tools, requiring significant computational skills. More recently, a new generation of interactive data wrangling tools instead uses visualization, interactive specification of rules, and machine learning to improve the efficiency and scale of data manipulation tasks while also providing accessibility to a broader set of analysts~\cite{kandel_wrangler:_2011, verborgh_using_2013, trifacta_trifacta_2012}.

%Even if available data is potentially useful to answer an analysis question, informal accounts put the effort required to wrangle and clean up a dataset as the most tedious aspect of a data analysis project~\cite{kandel_research_2011}. Data wrangling, defined as \textit{``iterative data exploration and transformation''}~\cite{kandel_research_2011} of datasets includes data cleanup, merging datasets, and data transformations. Although historically wrangling was done with scripting languages such as Python, Perl and R, or manipulation in spreadsheet tools, a new generation of interactive data wrangling tools~\cite{kandel_wrangler:_2011, verborgh_using_2013, trifacta_trifacta_2012} use visualization, interactive specification of rules, and machine learning to wrangle tabular datasets. 

These powerful, interactive data wrangling tools, however, ignore a data type that is increasingly important~\cite{behrisch_commercial_2018}: networks. Whereas some datasets inherently represent a network that exists in the physical world, such as the connections between neurons in a brain or roads between cities, many other datasets also benefit from a network representation during analysis. The influence of social connections on obesity rates~\cite{christakis_spread_2007}, the spread of information via a digital media platform~\cite{bail_combining_2016}, or the evolution of sticky feet of geckos~\cite{hagey_tempo_2017} are but a few examples.
%Networks (or graphs) capture relationships (links) between entities (nodes). 
%Networks can be ``natural'', such as the networks of neurons in a brain or roads between cities; but they can also capture more abstract relationships, such as friendships between individuals in a social network. 

In such cases, an analyst has one, or possibly many, mental models of the data as a network. However, source data rarely conforms to the way an analyst thinks about it. To model data as a network, analysts must wrangle the dataset, often starting with tabular or key-value data. 
Transforming data itself can lead to new hypotheses, and thus a new network representation of the data. 
Also, new tasks often necessitate new data abstractions~\cite{munzner_nested_2009}. It stands to reason that the ability to rapidly and easily transform network data can foster creative visualization solutions and simplify both exploration and communication of the key aspects of a dataset.

Existing network wrangling tools, most notably Ploceus and Orion~\cite{heer_orion:_2014,liu_ploceus:_2014}, focus on creating an initial network model, but no tools yet exist to iteratively and interactively reshape the network model itself \newText{with operations such as converting between nodes and edges~\cite{nielsen_abyss-explorer:_2009}}.
\oldText{Several wrangling operations, especially non-tabular reshaping operations}\newText{Other operations that leverage edges, such as connectivity-based filtering~\cite{bigelow_jacob's_2019}, can currently be performed only with code.}
Consequently, the steep learning curve of programming languages creates an unnecessary barrier that prevents many analysts from wrangling their own networks.

%the mental model an analyst has about a dataset is of a network, yet the data is available in tabular form. Network modeling tools, such as Orion~\cite{heer_orion:_2011, heer_orion:_2014} or Ploceus~\cite{liu_network-based_2011, liu_ploceus:_2014}, can introduce links based on attributes and hence create a network from a tabular dataset. Their power to re-shape a network, however is limited. 

In this paper, we introduce Origraph, our primary contribution (see Figure~\ref{fig:teaser}). Origraph is a visual, interactive network wrangling tool that \newText{implements an expanded set of wrangling operations that} allow analysts to model and reshape networks from input data in various forms. The goal of Origraph is to allow analysts to translate their data into the network representation that is most suited to answer their analysis questions \newText{and refine or revamp that representation as analysis questions change over time}. 

\newText{We also contribute a discussion of network wrangling operations, propose several new operations, and then classify them into a preliminary taxonomy. The elicitation of these operations is grounded in a literature analysis, an analysis of user needs in prior projects, eliciting missing capabilities in existing tools, and our own experience in wrangling network data. Origraph implements all operations we identified.} Operations that are unique to Origraph are concerned with introducing new nodes, edges, or attributes based on leveraging network structures and multivariate attributes simultaneously. For example, Origraph supports an operation to introduce edges between two nodes if these nodes are connected by a path with specific properties. 

Origraph includes visualizations of the data that support analysts in their reshaping and analysis process. Dedicated views communicate the state of the network, the attributes of the nodes and edges, and a sample of the network as currently modeled. Specialized views and algorithms support analysts in making network wrangling decisions. Origraph is web-based and open-source. A prototype is available at \url{https://origraph.github.io/}.

%Origraph is not a data analysis tool, but instead provides analysts with an understanding of the more abstract network representation.

We designed Origraph with a diverse audience in mind: from data journalists who analyze bot networks on social media, to social scientists who investigate the spread of specific terms in political circles, to biologists who study protein interaction. We do not expect users of Origraph to be able to program. Some advanced functionality is made available for more skilled analysts, such that they can write expressions slightly above the level of formulas in spreadsheet software to perform sophisticated filtering and aggregation. At the same time we believe that Origraph can also speed up the wrangling process of skilled programmers.

We validate Origraph in two complex network modeling use cases. First, we reshape a movie dataset so that we can investigate gender biases in recent popular movies. In the second use case, we integrate data from various sources and build a network that allows us to investigate how money from donors could influence votes in the US Senate on issues related to the war in Yemen.

\section{Related Work}

Origraph draws on related work in graph editing, tabular data wrangling, graph databases, and network modeling. Graph editing refers to tools aimed at visualizing and editing existing networks. Tabular data wrangling tools work with tabular data and excel at data transformation and querying. Graph databases are optimized for storing, indexing, and querying large networks. Network modeling involves extracting a network from tabular data. We discuss relevant prior work in the respective subsections below. \newText{Various network visualization systems support basic operations, such as aggregating nodes into supernodes, or filtering nodes and edges. We limit our discussion here to tools and techniques that support more sophisticated wrangling operations and refer to review articles~\cite{nobre_state_2019, mcgee_state_2019} for a survey of more general network visualization methods}.

\subsection{Graph Editing}

A wide array of tools are designed to allow users to visualize and edit networks and their associated attributes.  Tools such as Cytoscape~\cite{shannon_cytoscape:_2003} and Gephi~\cite{bastian_gephi_2009} focus on the topology of the graph and offer several editing features. Similarly, Graphviz~\cite{ellson_graphviz:_2001} is a collection of graph drawing tools, including layout programs and customizable graph editors. 

Tools in this space allow users to modify a network, mainly by creating or deleting nodes and edges. This level of editing is useful for tasks such as finding and correcting mistakes in the data or inputting new data. However, these edits are not based on rules; rather they are limited to the instance level and hence do not generalize to the entire network. Filters are a common exception to this restriction: most network visualization tools support filtering nodes or items.

Generally, graph editing tools assume a well-defined network as input. These tools are primarily designed to represent network models as they exist, and do not have features aimed at creating or deriving new models. The Tulip framework~\cite{auber_tulip_2017} touches on network modeling by enabling users to import data and generating multiple data models for users to visualize and explore. However, users do not have control over how these data models are generated, nor can they modify them. 

\subsection{Wrangling Tabular Data}

%Unlike graph editing tools, Origraph focuses on creating new network models and allows for easy reshaping of existing models. 
Data wrangling applications for tabular data include tools such as Google Refine~\cite{huynh_google_2011}, Data Wrangler~\cite{kandel_wrangler:_2011} and its commercial successor Trifacta Wrangler~\cite{trifacta_trifacta_2012}, and Microsoft Excel, which focus on data transformation and cleanup. These systems enable analysts to reformat input data to best suit their analysis tasks, but they are not designed to support network data. 

Some data wrangling tools use network visualization in the process of wrangling data. D-Dupe~\cite{bilgic_d-dupe:_2006} uses a network perspective to help resolve duplicate entities while cleaning a dataset. Schema Mapper~\cite{robertson_visualization_2005} uses network visualizations to explore how one hierarchical dataset maps to another. GraphCuisine uses a visual interface to generate random networks~\cite{bach_interactive_2013}. Although these approaches utilize network visualization in the wrangling process, and support wrangling tasks on non-tabular data, they do not wrangle networks themselves.

NodeXL~\cite{smith_analyzing_2009}, an extension to Excel, allows users to import, visualize, and transform network data. However, NodeXL provides only a minimal set of network transformation features.
%and a more expressive set of operations would best be achieved by enabling interoperability with other libraries. 
Moreover, Liu \textit{et al.}~\cite{liu_ploceus:_2014} point out that conducting extensive network reshaping with NodeXL would require users to be Excel experts.

\subsection{Scripting for Graph Wrangling}

Another domain of related work is network-specific technologies ranging from libraries such as NetworkX~\cite{hagberg_exploring_2008} and mully~\cite{hammoud_mully:_2018} to graph databases such as neo4j~\cite{neo4jinc._neo4j_2010}, OrientDB~\cite{orientdb_orientdb_2010}, and GraphDB~\cite{ontotext_graphdb_2000}.
%\oldText{These systems are optimized for storing, indexing, and querying large networks. Specific languages have been developed to query and manipulate graphs in these databases.}
Similar to Origraph, these approaches allow users to create and transform network models as needed. However, creating or changing a network model in such systems must be done through scripting, requiring users to be proficient in the context-specific language. In contrast, Origraph does not require programming for most operations and allows users to create and transform network models with an interactive interface.
 
\subsection{Network Modeling}

Network modeling, i.e., the concept of creating networks from tabular data, has been explored by several tools. The need for these tools arises from the frequent storage of network data as tables, containing a list of items and their associated attributes. These tools commonly also provide visualization to better understand and explore the resulting network structure. 

Commercial systems such as TouchGraph Navigator~\cite{touchgraph_graph_2018} and Centrifuge~\cite{centrifugesystems_centrifuge_2018} support network modeling by allowing users to create attribute relationship graphs from tabular data. Attribute relationship graphs, which were introduced by Weaver~\cite{weaver_multidimensional_2010}, refer to graphs where attributes are connected based on co-occurrence. %The authors distinguish between attribute relationship graphs and attributed graphs, where an object is connected to its attributes. 
A related approach, used by both PivotGraph~\cite{wattenberg_visual_2006} and HoneyComb~\cite{vanham_honeycomb:_2009}, generates new network models by aggregating nodes that have a certain attribute.%, an operation called `rolling up'.

The two systems most related to Origraph are Orion~\cite{heer_orion:_2014} and Ploceus~\cite{liu_ploceus:_2014}. Both systems model attribute relationship graphs based on tables from relational databases. Ploceus's approach is based on relational algebra whereas Orion uses relational tables as its base and represents networks as edge tables. Although Orion and Ploceus are well suited for creating networks from tabular data, they do not support reshaping existing networks. The ability to iterate on chosen data abstractions is restricted to creating a new network model from the raw data. Each distinct network model needs to be specified from the ground up; users cannot create alternative network models with an existing model as a starting point. Both Orion and Ploceus do support node attributes, but unlike Origraph, do not support edge attributes beyond simple edge weights. We discuss differences and similarities between Origraph and these two tools in more detail in Section~\ref{sec:discussion}. %For a comparison of supported operations refer to  Table~\ref{table:operations}.
%Notably, both tools emphasize analysis of network data within the tool, whereas Origraph provides build-in visualization features, but also acknowledges that is designed as a wrangling-first application, with visualization for analysis mostly used for validating wrangling operations.

Another related modeling tool is Graphiti~\cite{srinivasan_graphiti:_2018}, which uses demonstration-based interaction to create edges in networks. Graphiti infers possible data abstractions from instance-level operations. Although Graphiti supports modifying existing networks by adding new edge types, it does not support more extensive reshaping operations such as changing nodes to edges or deriving new classes from attributes of connected nodes or edges. %Additionally, Graphiti is only suited to working with undirected, unipartite graphs. Origraph offers support for unlimited types of nodes and edges as well as directed edges, allowing users to connect data from multiple sources to generate a suitable network model.

Ultimately, Origraph builds on existing work by extending the concept of network modeling and providing powerful network reshaping features. In this vein, Origraph offers support for edge attributes, more expressive network modeling features such as faceting and deriving reduced attributes, and annotation features. Our work differs from other systems by giving users the ability to leverage existing network models to create more expressive abstractions for node and edges that best reflect their semantic understanding of their data\newText{, and by enabling rich edge semantics that other systems cannot represent}. %Additionally, unlike most previous efforts, Origraph is available as an open-source web application.

\section{Terminology}

Here we define terminology that we use to describe the operations, as well as the functionality of Origraph. In this work, we focus on modeling \textbf{networks} using \textbf{nodes} and \textbf{edges}. Both nodes and edges can have \textbf{attributes} associated with them. A \textbf{class} defines a common set of attributes for either nodes or edges: a \textbf{node class} defines a class of nodes, and an \textbf{edge class} defines a class of edges. We treat nodes and edges as fluid concepts that can change, and use the term \textbf{class} to generically refer to node and edge classes and \textbf{items} to refer to generic instances of nodes or edges. \textbf{Supernodes} are nodes representing a set of other nodes.

We use the term \textbf{network model} for the set of classes and their relationships. The network model describes relationships between types of nodes and edges on an abstract level and is independent from concrete instances of the classes, similar to how a database schema is independent from the specific data in the database. Concrete nodes and links make up the \textbf{network topology}. It is noteworthy that networks with trivial models (e.g., a single node class and a single edge class) can be arbitrarily large and complex.

Edges can be directed or undirected. Node and edge attributes can be numerical, ordinal, nominal, sets, or labels/identifiers. Attributes may also contain complex data structures such as nested hierarchies, lists, and objects.

\section{\newText{Eliciting Operations}}

\newText{We developed our set of proposed operations through a reflective analysis process, in which there is a direct mapping from user tasks to operations. Connecting nodes, for example, is both a user task and a network wrangling operation. We use the term operation, as we frame operations from the perspective of what a network wrangling system should support. In our elicitation, we reflected upon published task lists for visual graph analysis~\cite{lee_task_2006, pretorius_tasks_2014, nobre_state_2019}; related systems that support limited aspects of graph wrangling~\cite{liu_network-based_2011, heer_orion:_2011, bigelow_jacob's_2019} our own experiences working with domain experts on graph analysis~\cite{meyer_pathline:_2010, partl_enroute:_2012, partl_enroute:_2013, lex_entourage:_2013, partl_pathfinder:_2016, kerzner_graffinity:_2017, nobre_lineage:_2019, nobre_juniper:_2019, bigelow_jacob's_2019, nobre_state_2019} and those of colleagues~\cite[ for example]{wattenberg_visual_2006, elmqvist_zame:_2008, pretorius_visual_2008, nielsen_abyss-explorer:_2009}; and an informal task analysis that we conducted with new collaborators~\cite{miriahmeyer_collaborative_2018, bail_combining_2016, bail_cultural_2016, morrison_network_2017, lauritzen_rod-cone_2019}. Through an iterative, dialogical approach among the co-authors that involved discussions, written summaries, and design sketches, we considered the myriad of literature and experiential sources in determining the range of operations for wrangling and reshaping networks that are, or could be, useful for analysts. The final list of operations represents our proposed vision of what should constitute a feature-rich system for fully supporting a flexible approach to using networks.}

\newText{These operations could support a new approach to answering complex question like those we pose in Section~\ref{sec:use-cases}, one that makes use of networks as a data representation that can be derived from various input data sources and flexibly reshaped as an analysis progresses and new questions come up.} 

\newText{As interactive network wrangling is a relatively new research area, the ambition of our work is to expand the vocabulary of operations, and to classify these operations in our preliminary taxonomy.  
Important future work will be to identify a formal taxonomy and to study which operations are useful for specific tasks and user group combinations.}

\section{Operations}
\label{sec:operation}

We classify \newText{the operations we identified into three categories, which constitute our taxonomy:} \textbf{modeling operations} that modify the network model and network topology by introducing new node and/or edge classes in a network; \textbf{item operations} that modify the network topology by removing or introducing items (instances of links or nodes); and \textbf{attribute operations} that manipulate the attributes of, or add new attributes to, existing classes. 

\newText{We illustrate these operations with a simple movie network consisting of movies, actors, and roles with attributes such as genre, company, and gender. We introduce a more complex wrangling workflow with a real movie dataset in Section~\ref{sec:us-movies}. We provide a comparison of which operations are supported in other systems, and which are unique to Origraph in Supplementary Table 1.}%, iterating through several different network models such as the one in Figure~\ref{fig:teaser}.}

Although we do not claim that this list is exhaustive, we demonstrate that the combination of these operations extends the space of transformations currently possible with Orion and Ploceus~\cite{liu_ploceus:_2014, heer_orion:_2014}.

%Note that all of these operations are \textbf{rule based}. That means that we exclude operations executed on specific instances, such as adding an edge or deleting a node based on an analyst's knowledge. Instead, we can delete edges or generate nodes based on a rule that can be applied to the whole network. For example, we could introduce edges between actors if they were born in the same year.

%It should also be noted that none of these operations fundamentally requires programming ability. Although each individual operation has a precedent in programming languages, and in some cases, in existing network modeling tools, Origraph demonstrates how a broader set of operations can be performed interactively.

\subsection{Modeling Operations}

Modeling operations affect the network model by introducing or removing node or edge classes. Modeling operations also affect the network topology because they implicitly create new node and edge instances.

\addIcon{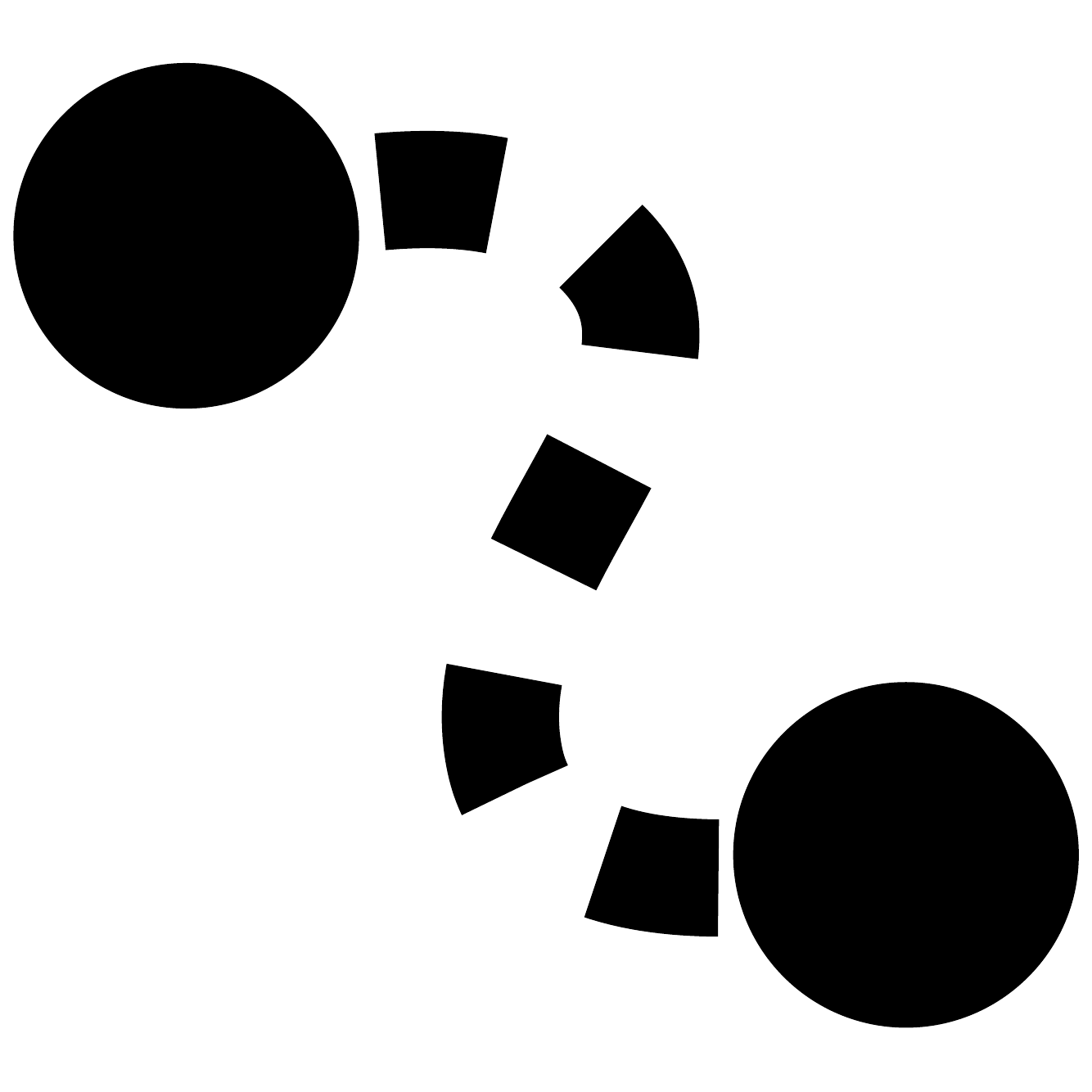} \textbf{Connecting or disconnecting} items is the most fundamental modeling operation. Connections can be established by leveraging the primary key / foreign key approach well known in the database literature, where each item of a tabular dataset has a unique ID, and another column has a foreign key pointing to the primary key of the same or another table. This is also a common way to store graphs in non-volatile memory: many network file formats store lists of nodes and lists of links between these nodes. We also consider cases where an item in a table stores a list of connection targets using foreign keys. More generally, connections can be derived from arbitrary attributes, for example by (partially) matching strings, or by evaluating arbitrary functions on attributes. In the movie dataset, for example, we could introduce edges between movies that have a significant number of female actors to form a clique of gender-balanced movies.

\begin{figure}[h]
  \centering
  \vspace{-4mm}
  \includegraphics[width=\linewidth]{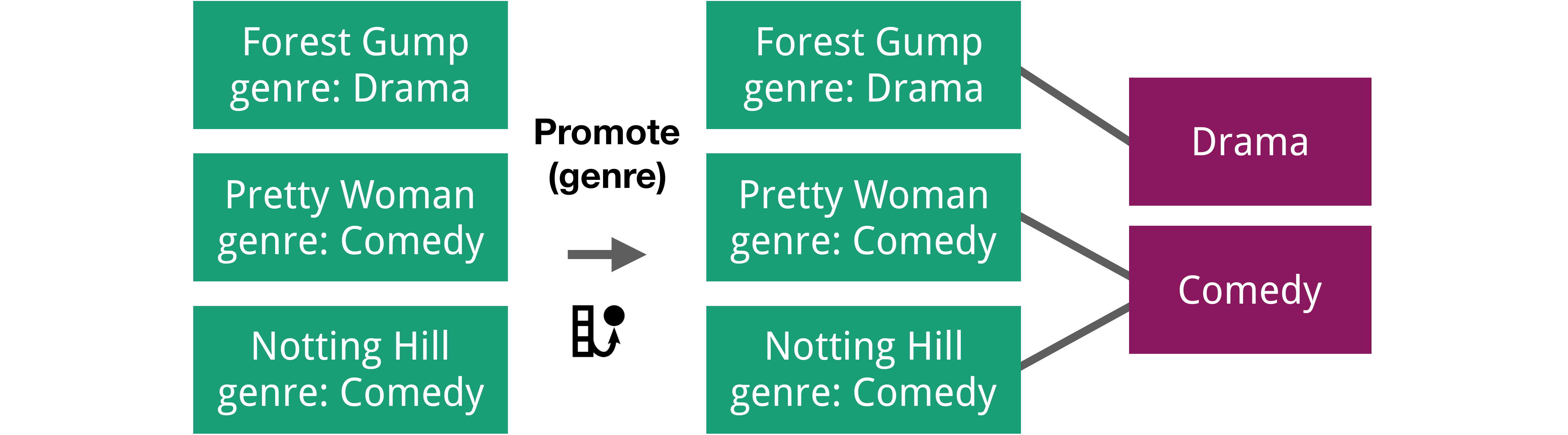}
  \vspace{-6mm}
  \caption{Promoting the genre attribute of the three movies. A new genre node class with two instances contains two nodes with the unique genres. The new nodes are connected to their source nodes.}
  \vspace{-3mm}
  \label{fig:op_promote}
\end{figure}

\addIcon{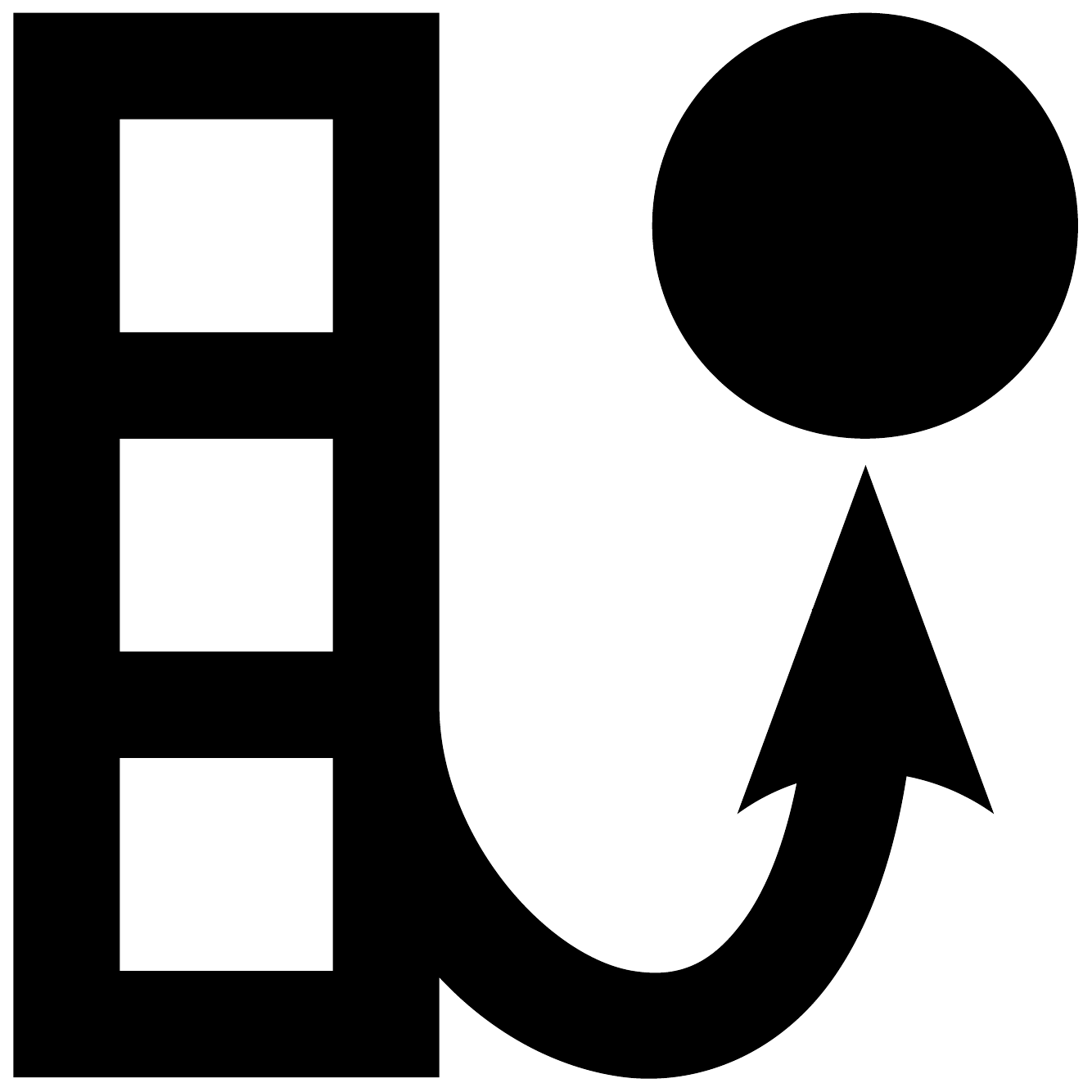} \textbf{Promoting attributes} allows users to promote the unique values of an attribute column to a new class (Figure~\ref{fig:op_promote}). In the movie example, we could take a column containing film genres and promote these to a separate ``genres'' class, while at the same time introducing edges between the new genre nodes and the movie nodes. 

%An important consideration when promoting unique attributes to their own classes is what to do with the other attributes of source class. The appropriate solution, such as dropping all attributes, averaging, summing, or storing as nested lists, will depend on the data and task. For example, when promoting genres, it is reasonable to also store the number of movies and the sum of the box office income for each genre. 

\begin{figure}[h]
  \centering
    \vspace{-3mm}
    \includegraphics[width=\linewidth]{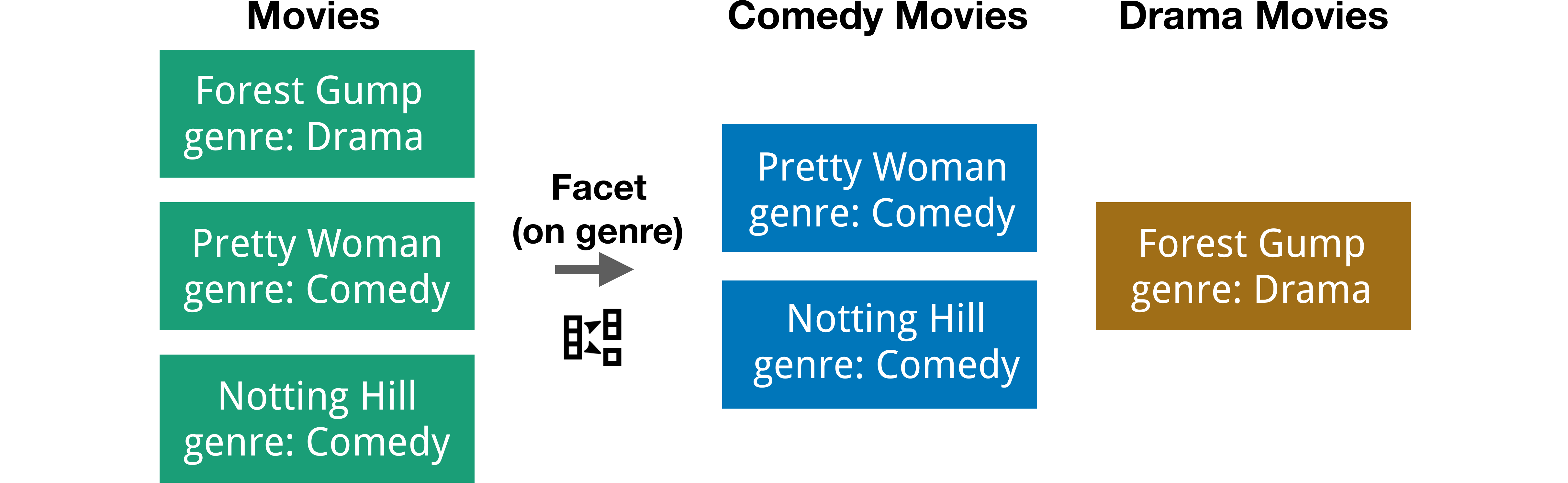}
      \vspace{-6mm}
  \caption{Faceting a movies node class based on genres. The facet operation results in two new node classes, one for ``Comedy Movies'' and the other for ``Drama Movies''.}
    \vspace{-3mm}
    \label{fig:op_facet}
\end{figure}

\addIcon{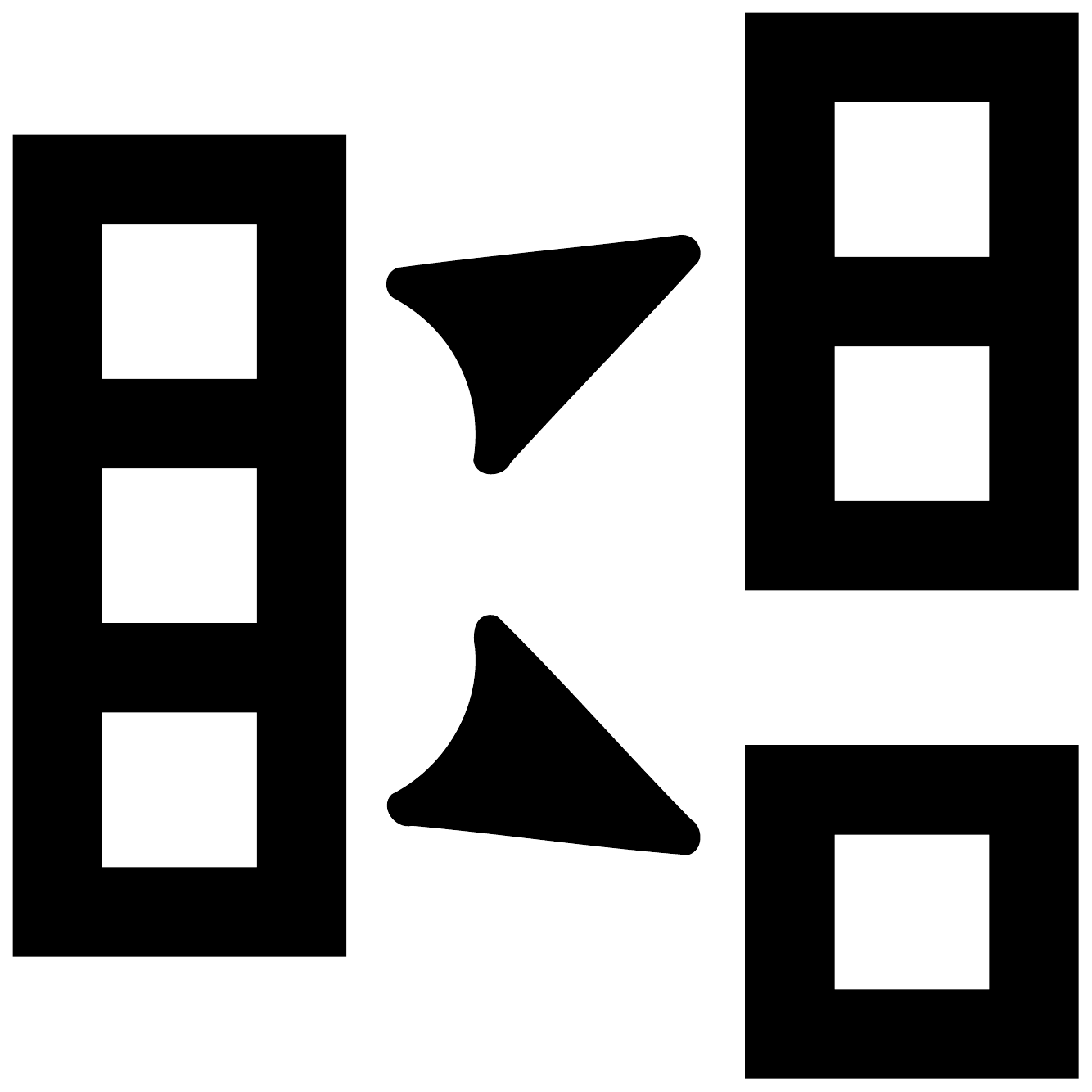} \textbf{Faceting} slices a class based on the value of an attribute and creates new classes for each slice, as illustrated in Figure~\ref{fig:op_facet}. An example is to facet the movie class on the genre attribute, which generates a new class for each unique genre containing only the movies of that genre. %Faceting is often a precursory step to other operations. In the genre case, for example, we could subsequently aggregate selected genres into supernodes.

\begin{figure}[h]
  \centering
      \vspace{-3mm}
    \includegraphics[width=\linewidth]{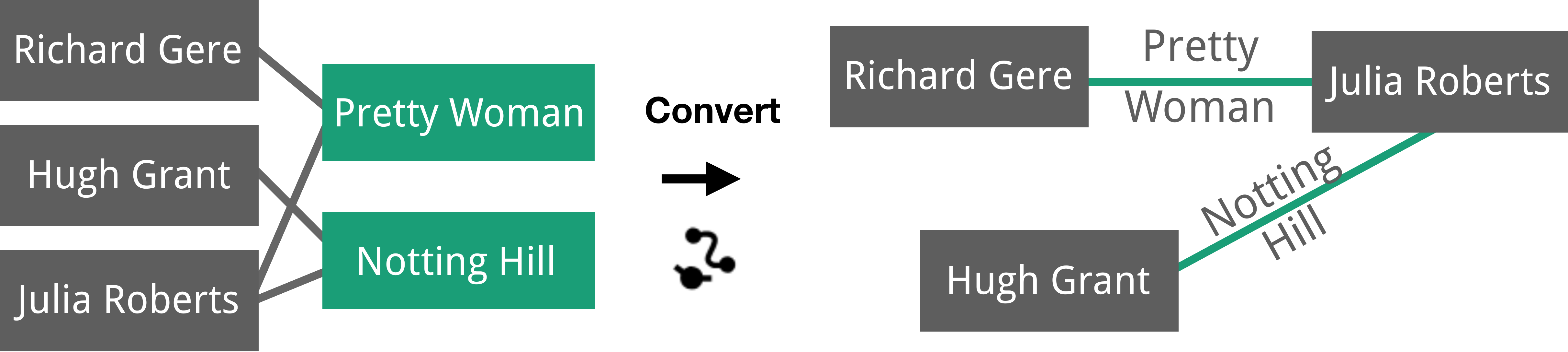}
     \vspace{-6mm}
  \caption{Converting nodes to edges. Here, movie nodes are converted to edges, resulting in a network linking co-actors.}
     \vspace{-3mm}
    \label{fig:op_convert_node_to_edge}
\end{figure}

\addIcon{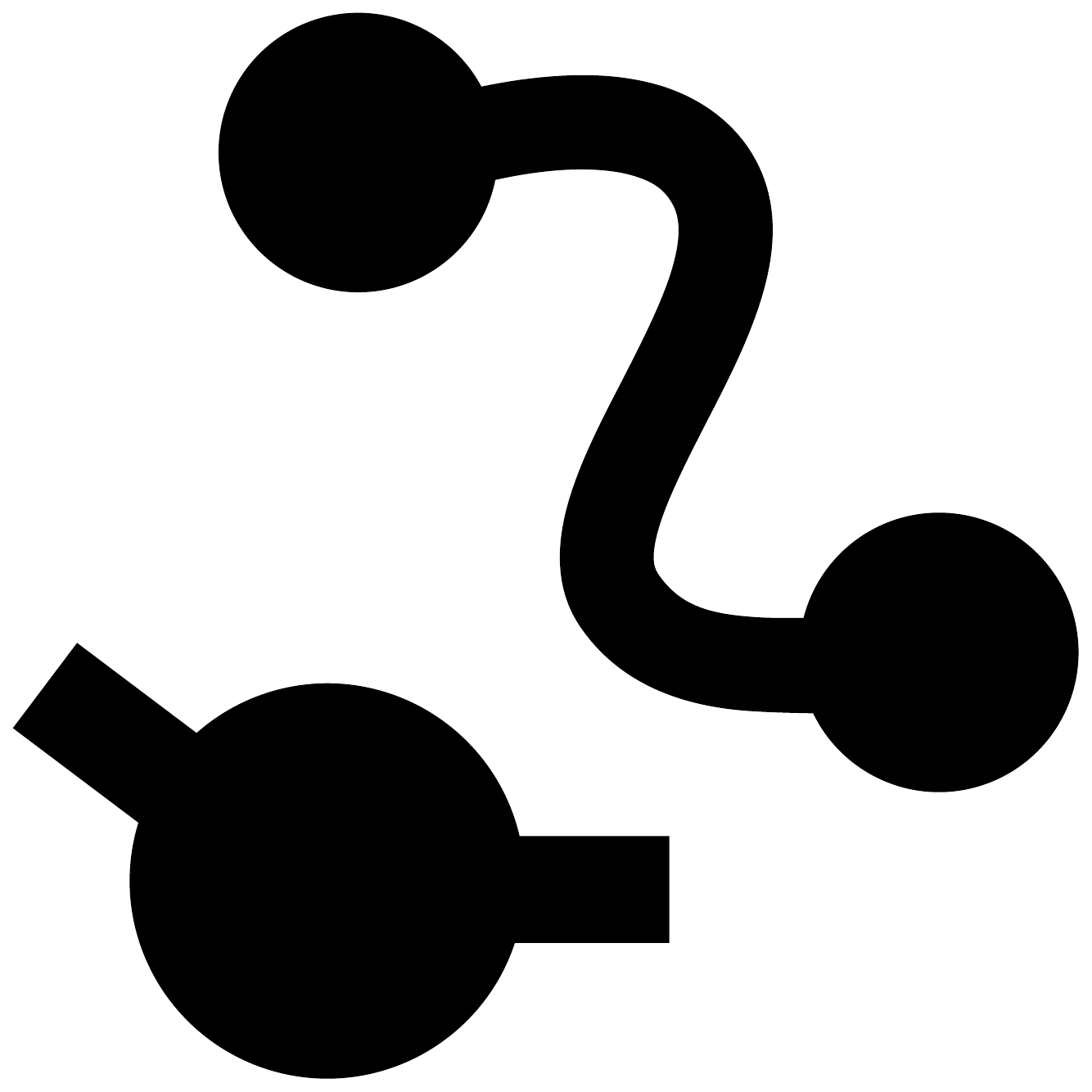} \textbf{Converting between nodes and edges} transforms connected nodes into edges, or \textit{vice versa}, retaining the connectivity and semantics of the network, as illustrated in Figure \ref{fig:op_convert_node_to_edge}. For example, given actors connected to movies, we convert the movie nodes to edges, which results in a collaboration network between actors, where edges connect actors who have acted together in a movie. 

\begin{figure}[h!]
  \centering
        \vspace{-3mm}
    \includegraphics[width=\linewidth]{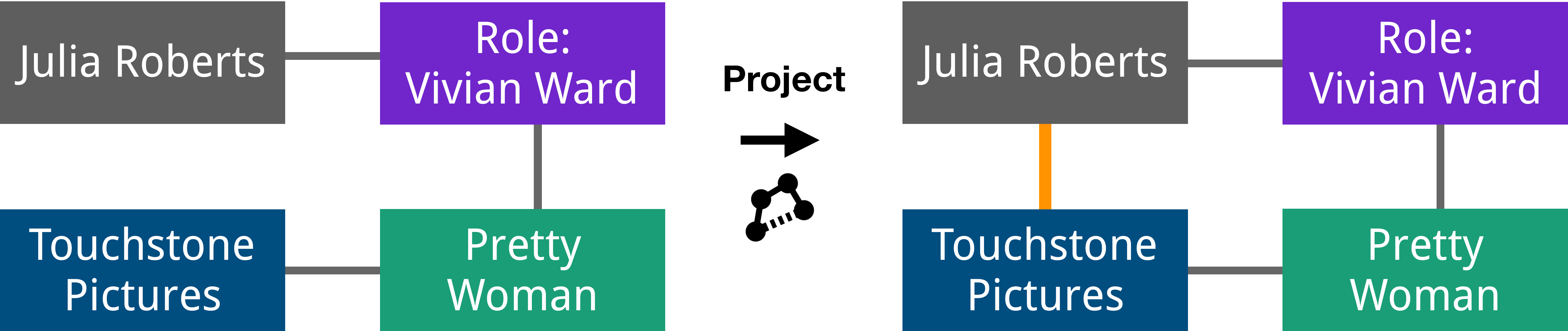}
       \vspace{-6mm}
  \caption{Projecting an edge. Paths connecting an actor to a role to a movie to a production company are projected to an actor--production-company edge (orange). }
       \vspace{-2mm}
    \label{fig:op_project_edges}
\end{figure}

\addIcon{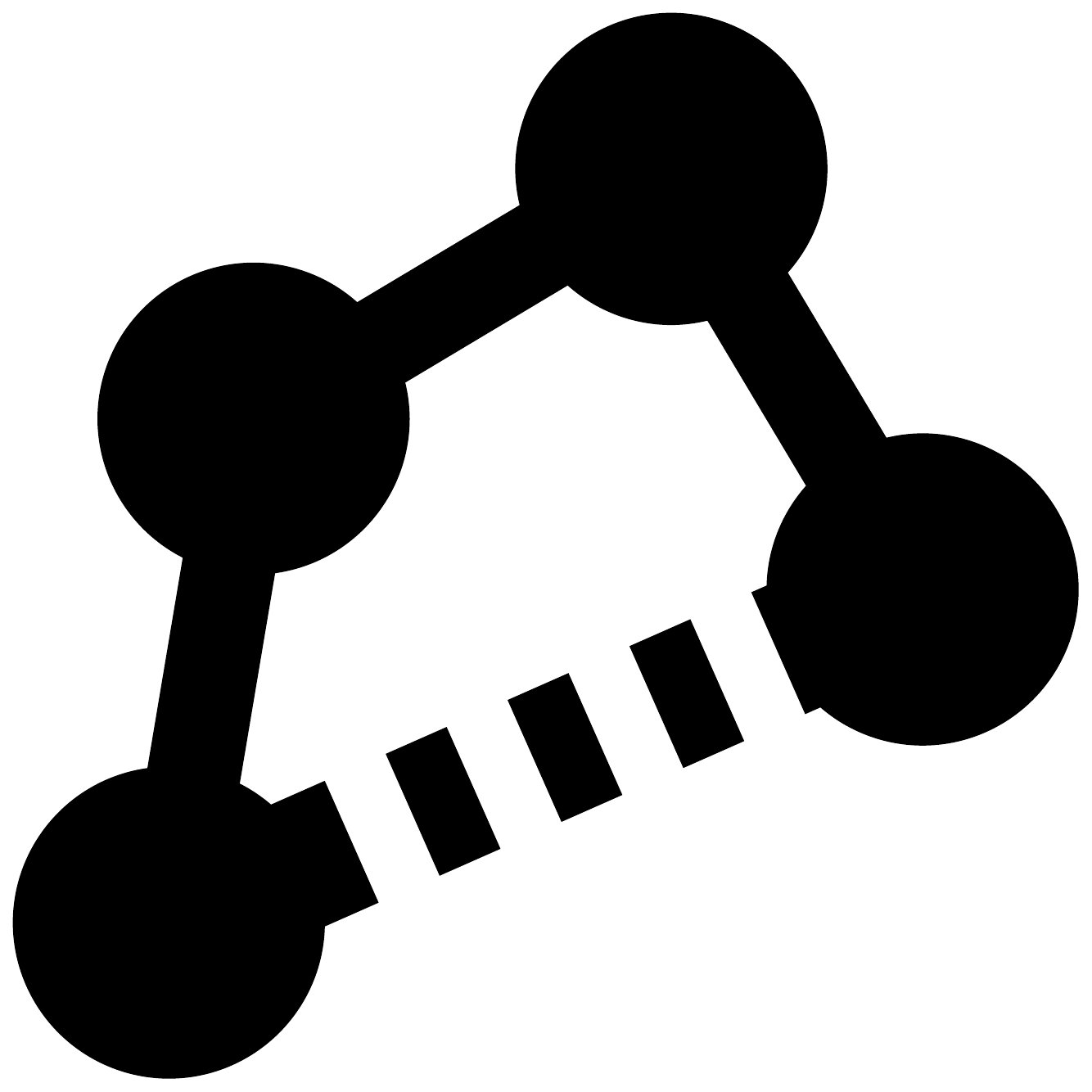} \textbf{Edge projection} introduces an edge based on a path in the network model between nodes, as illustrated in Figure~\ref{fig:op_project_edges}. The path can be specified with a set of rules leveraging the classes and the attributes of the network. For example, in an actor--role--movie--production-company network, edge projection can be used to connect an actor with the production-company. Another example is to project edges from an actor--movie--actor relationship to an actor--actor relationship, limiting edges to financially successful collaborations by considering only edges that transit through movies that had a box office return above a specified number.

\begin{figure}[h]
  \centering
          \vspace{-2mm}
    \includegraphics[width=\linewidth]{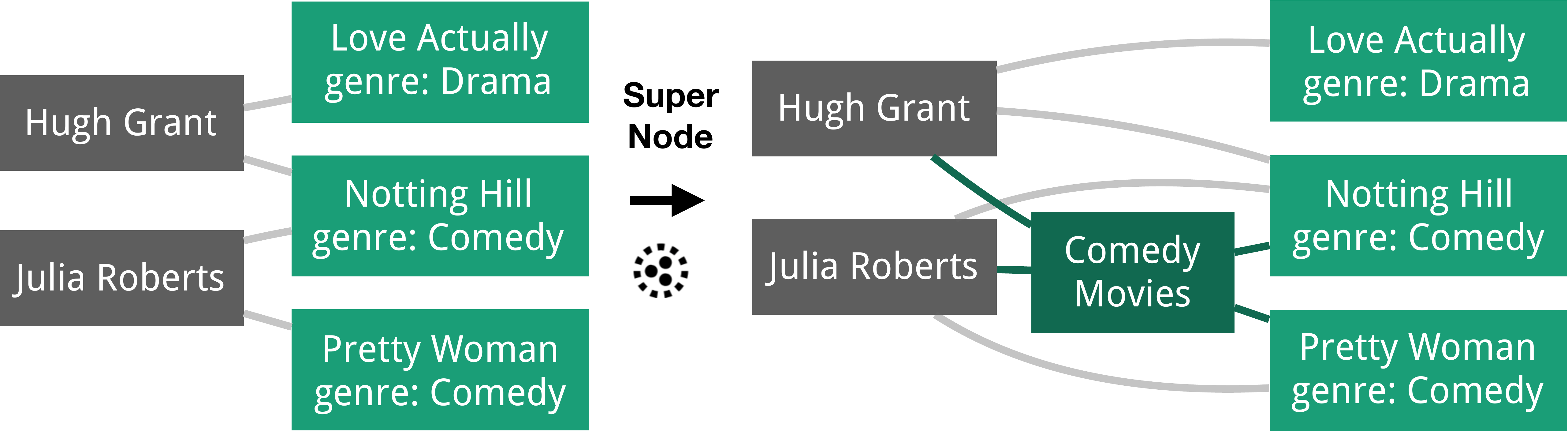}
        \vspace{-6mm}
  \caption{Creating a supernode. Notting Hill and Pretty Woman are combined into the supernode Comedy Movies. This supernode inherits all edges and also has edges to the aggregated movies.}
      \vspace{-1mm}
    \label{fig:op_supernodes}
\end{figure}

\addIcon{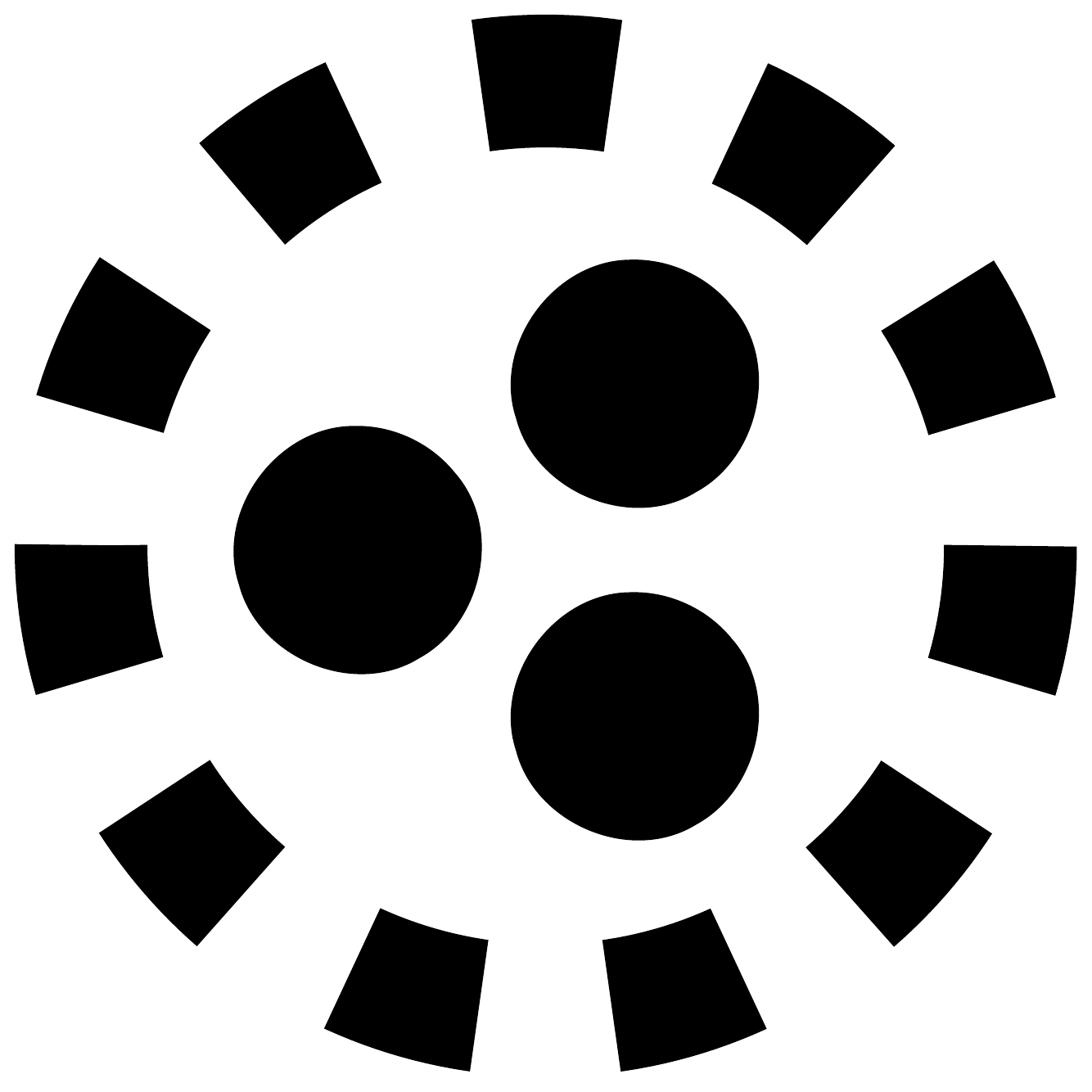} \textbf{Creating supernodes} aggregates the information of multiple nodes into a single supernode that represents all the constituting nodes. For example, we could create a supernode that represents all comedy movies (Figure \ref{fig:op_supernodes}). %When creating supernodes, we can also apply an aggregation function on attributes, such as summing or averaging the attributes of constituting nodes. 
There are different ways to realize a ``create supernode'' operation: one can retain all aggregated nodes and edges, or replace the aggregated nodes with the supernode. In Origraph, we retain all nodes and edges, which can be filtered out in a subsequent step. 

\addIcon{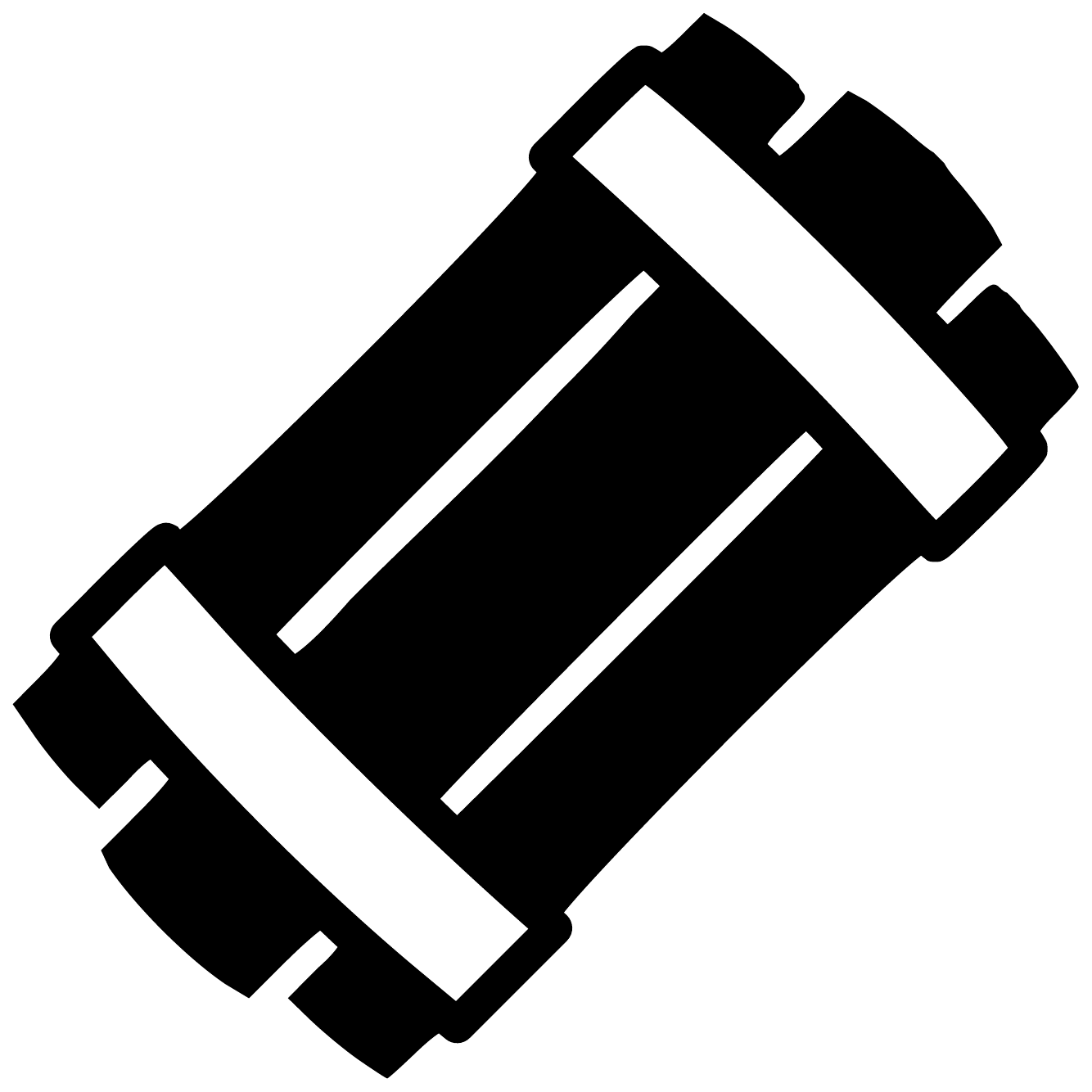} \textbf{Rolling up edges} combines parallel edges, or multiple edges that connect the same two nodes, into a single edge of a new edge type. For example, in a co-appearance network of actors, some actors might have appeared together in multiple movies, represented by multiple edges. The rollup operation results in a single new edge connecting these actors. %Again, aggregation functions for attributes could be applied. For example, if an actor plays more than one role in a movie, and therefore is connected to the movie by more than one edge, rolling up these edges would result in a single edge.

%\subsection{Other Wrangling Operations}

%\mm{Need intro that says why these are important. Logic from AL: although wrangling the model is the most important contribution of our paper, I think that the others (items and attributes) are also important and often necessary for subsequent model wrangling.}

\subsection{Item Operations}
Item operations change the number of items in existing classes. They may leverage the network model, but they do not modify it. Item operations affect the topology of the network, as they manipulate which nodes and edges exist.

\addIcon{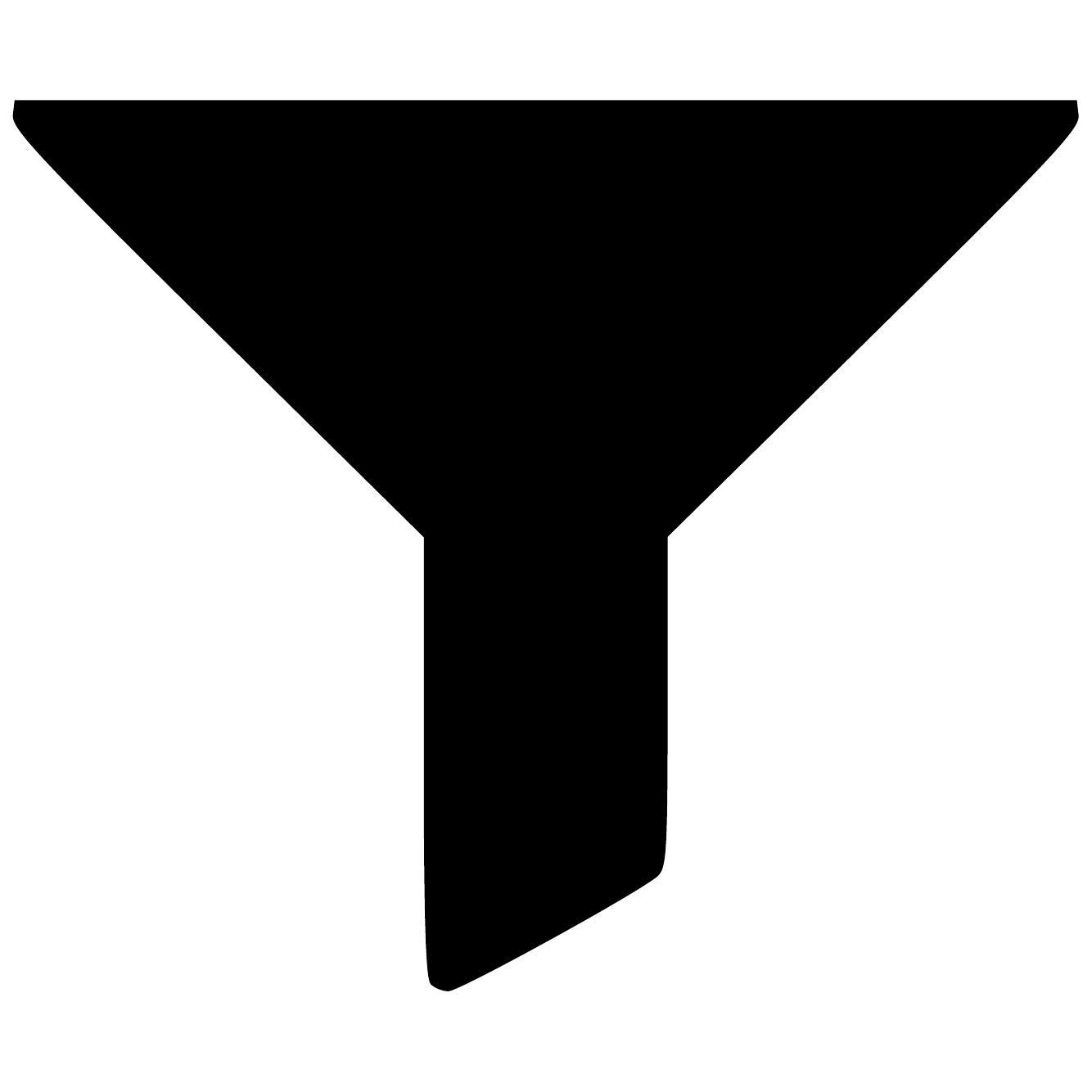} \textbf{Filtering by attributes} 
removes nodes or edges based on the values of an attribute. For example, we could filter by removing all movie nodes that grossed less than \$10 million, or remove all actor-movie edges where the role was not a speaking role. 

% One of the main goals of Origraph is producing a dataset that is suited to the analyst's specific questions. Aside from the larger graph reshaping features, this often involves filtering the initial data down to only a relevant subset.  Origraph allows user to filter based on any class attribute, leading to a dataset tailored to users' specific needs.

\addIcon{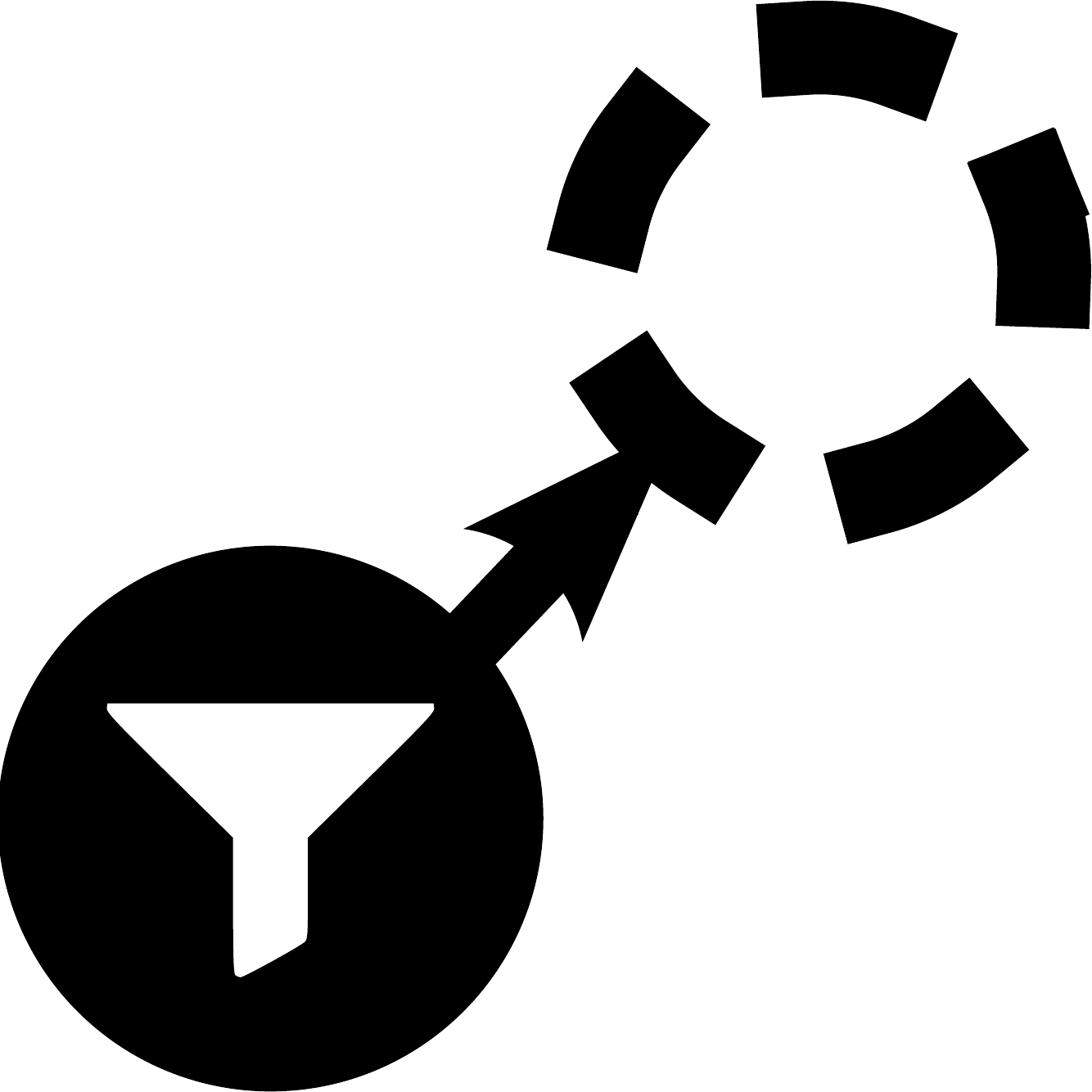} \textbf{Connectivity-based filtering} describes filter operations on items (nodes or edges) that leverage the connectivity of a network. In the movie network, where actors are connected to movies they acted in, an example of connective filtering is to remove all actors who have never acted in a movie that grossed more than \$100 million. Connectivity-based filtering can also leverage complex, multi-hop operations~\cite{bigelow_jacob's_2019}.

% (filtering based on connected node / edge topology, or indirect attributes

\subsection{Attribute Operations}
Attribute operations modify class attributes or create new ones, but they do not impact the network model or the network topology. They are, however, an important prior or subsequent step in many modeling operations. %For example, modifying an attribute of a node in preparation for a \connect connect operation to another node is frequently necessary, as is inheriting attributes from constituting nodes after a \supernode create supernode operation.

\addIcon{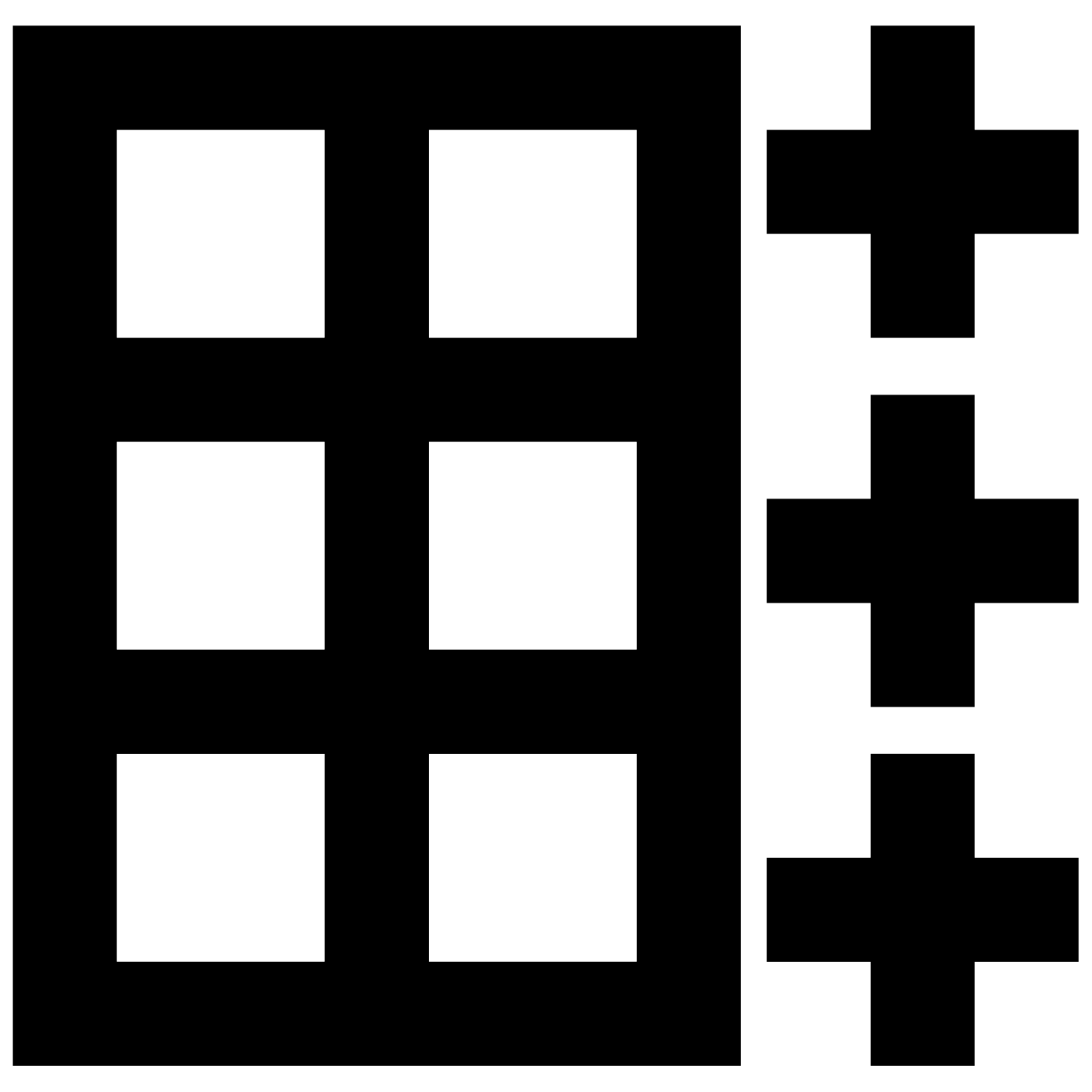} \textbf{Deriving in-class attributes} leverages existing attributes in a node or edge class to derive new ones. For example, for actor nodes with birth and death year, we could derive the attribute ``age'', representing either the actor's current age, if they are alive, or their age at death.

% Once a dataset has been loaded into Origraph, users may wish to compute new attributes, based on exisiting ones. Origraph supports attribute derivation within a given node or edge class where users can leverage existing attributes to derive a new class attribute. For example, if users have birth and death year as attributes in a Person class and want to compute an age, they can subtract death year from birth year and create a new age attribute on that class.

\begin{figure}[htbp]
  \centering
        \vspace{-2mm}
    \includegraphics[width=\linewidth]{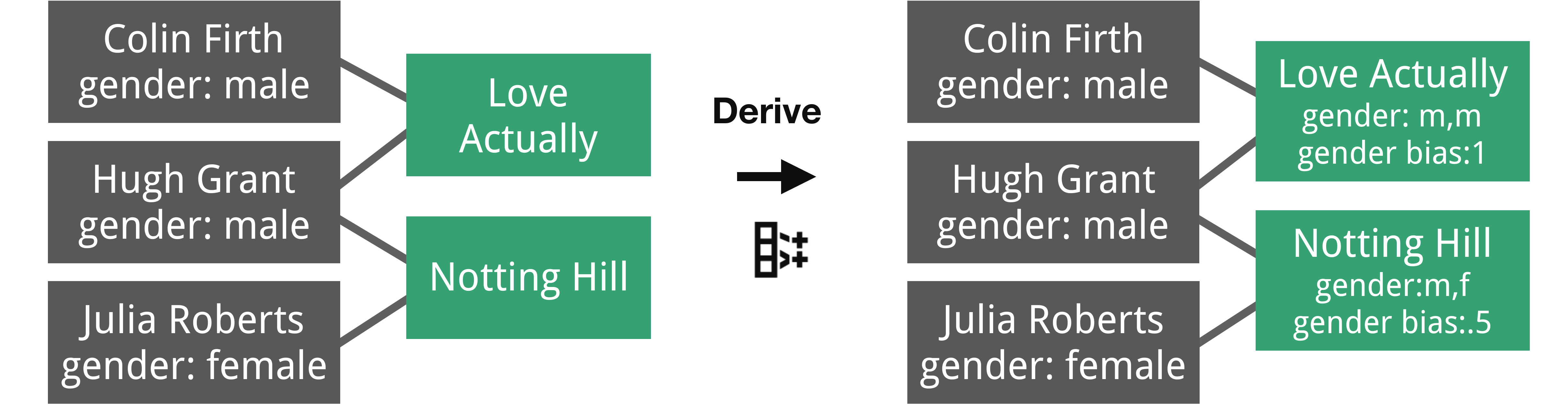}
          \vspace{-6mm}
  \caption{Deriving attributes across connected classes. Nodes of the movie class derive attributes from the adjacent actor class. In this case, the gender of the actors is used to calculate a gender bias score.}
        \vspace{-1mm}
    \label{fig:op_reduce_attributes}
\end{figure}

\addIcon{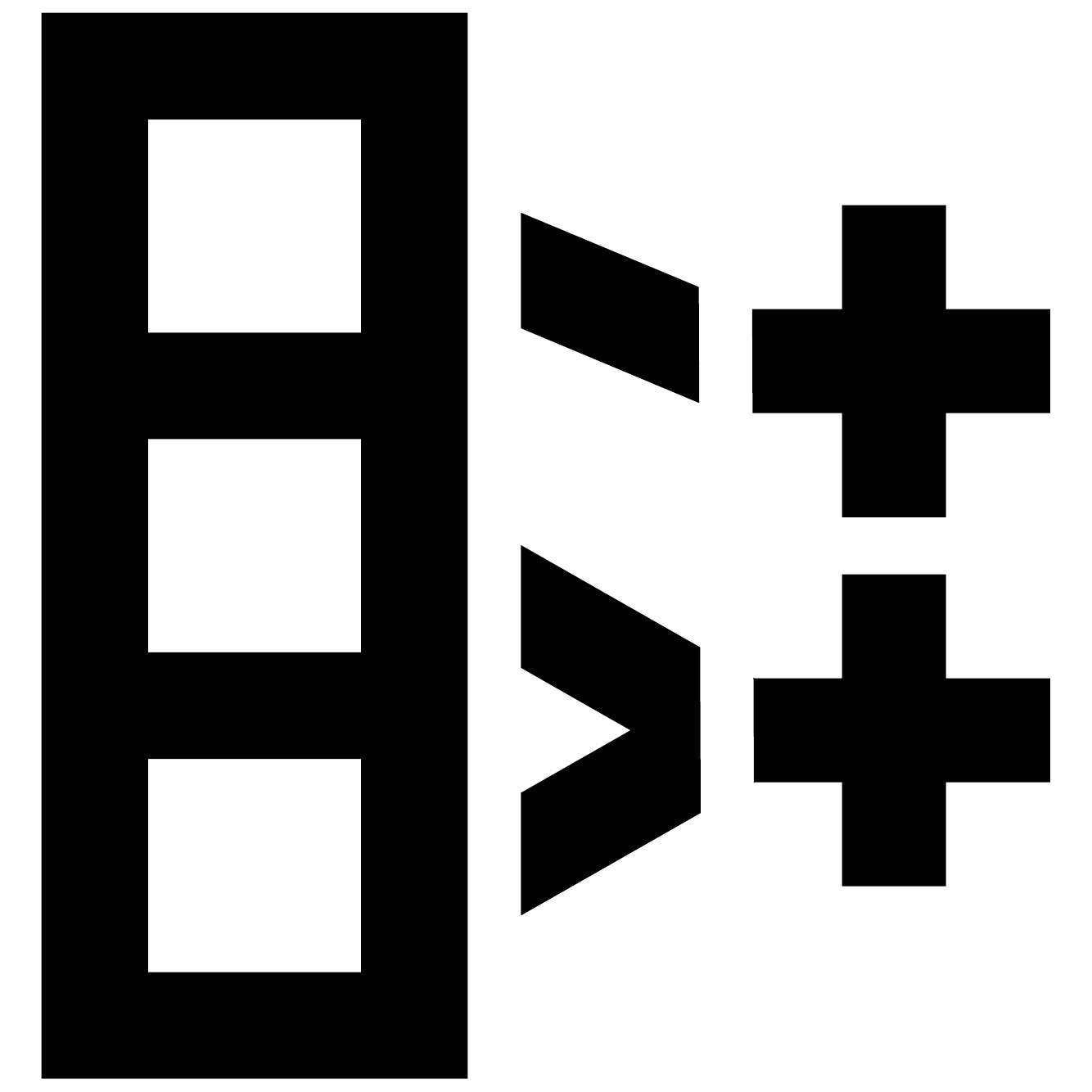} \textbf{Connectivity-based attribute derivation} is concerned with deriving attributes for a node or edge class based on attributes in a possibly indirectly connected class (Figure~\ref{fig:op_reduce_attributes}). As an example, in an actor-movie network, we can compute a new attribute ``gender bias'' on the movie class by iterating through all the actors connected to each movie and dividing the number of men by the total number of actors in that movie. %Because this operation generates a single value for every movie, users must specify to how to reduce all the connected elements---in this case actors---to a single value.
Connectivity-based attribute derivation can be very useful as a follow-up step to many modeling operations. For example, when \supernode creating supernodes, it is often relevant to also aggregate some of the attributes of the original nodes into the supernode. Combined with modeling operations such as \promote promoting attributes, \edgeProjection edge projection, \supernode creating supernodes, or \rollup rolling up edges, this is the network version of the split-apply-combine (or map-reduce) strategy common in tabular data analysis~\cite{wickham_split-apply-combine_2011}. As with tabular data, potentially relevant apply functions are immensely diverse.  

\addIcon{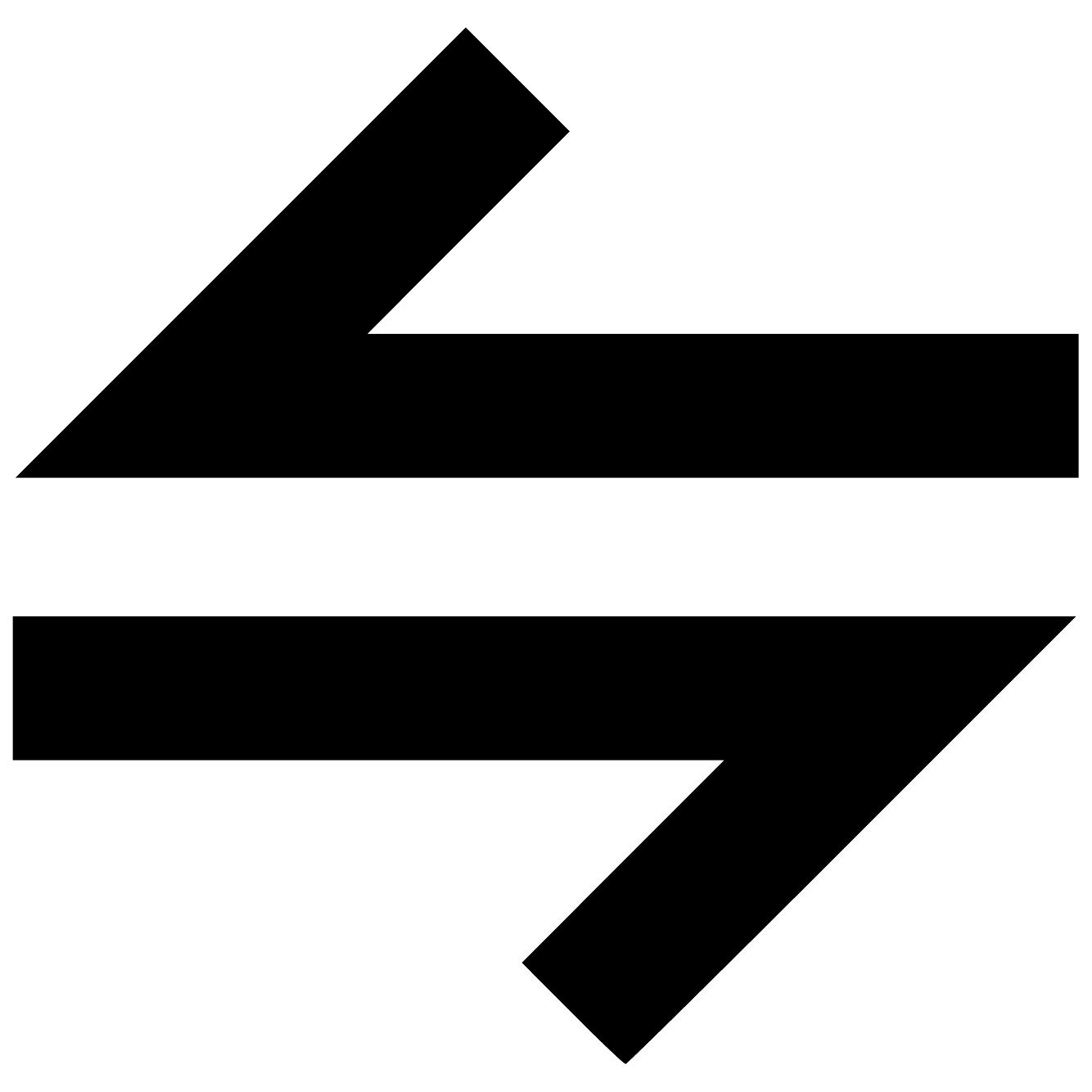} \textbf{Changing edge directionality} introduces directionality to, or removes it from, previously undirected edges, or changes the direction of edges. Like all other operations, changing edge direction is rule based. For example, in a network of movies and actors, edges between movies and actors can be made directional to represent an ``acted in'' relationship.

% Similar to the operation of deriving adjacent attributes, Origraph allows user to leverage existing attributes on connected classes to derive new ones. 

\subsection{Housekeeping Operations}

In addition to these data wrangling operations, a system implementing these operations will also need to enable a series of basic operations, such as importing data from various file formats, extracting nested file structures, exporting data for consumption by graph visualization tools, renaming classes and attributes, removing or hiding classes or attributes, etc. Since these are not specific to network wrangling, we do not discuss them in detail.

\subsection{Discussion}

%We explicitly exclude an operation for creating ``superedges'' by aggregating edges, because this operation would result in hyperedges, which lack widespread support in graph visualization tools.

\newText{
Several of these operations are available in prior systems, yet three operations are unique to Origraph. We discuss the merits of these novel operations here. \convert \textbf{Converting between nodes and edges} is a critical operation when working with multi-typed networks. Consider the case of movies and actors. In the source network, only relationships between movies and actors are captured. If analysts desire to investigate collaboration patterns, they have to convert the movie nodes to edges, capturing collaboration between actors.  \deriveConnectedAttribute  \textbf{Connectivity-based attribute derivation}, in contrast, is critical for analyzing network effects in datasets, i.e., effects that propagate along connections, a question that is fundamental to, e.g., studying social networks~\cite{christakis_spread_2007}. Deriving attributes based on distant connections makes it possible to reveal these network effects at the nodes they influence. For example, we can identify gender bias of movies by deriving attributes from the actors, or, even further removed, create a gender bias score for production companies based on attributes of nodes (actors) that are twice removed. \connectiveFilter \textbf{Connectivity-based filtering}, in turn, is a logical extension of attribute derivation: it allows analysts to leverage network effects at arbitrary distances to simplify the network to suit their analysis needs.
}

Some of the described operations can be achieved by sequentially executing other operations. For example, \connectiveFilter~\textbf{connectivity-based filtering} can be achieved by first \deriveConnectedAttribute~\textbf{deriving} attributes based on connectivity and then \filter~\textbf{filtering} based on attributes. Similarly, \convert converting nodes to edges could be achieved by first \edgeProjection~\textbf{projecting edges} and then deleting the intermediate nodes and edges. We have chosen to include these operations nevertheless, because we believe that they are closer to an analyst's mental model and require less indirection.% An implementation of these operations would certainly leverage the more basic operations.  

\section{Origraph Design}

We implemented the network wrangling operations we identified in Origraph, complementing them with visualizations supporting the operations, showing the state of the network in various views. A key design goal of Origraph is to immediately visualize the effect of an operation on the network model, the underlying topology, and the attributes, so that analysts can easily understand their actions.

%A design principle Origraph follows is \textit{in-place manipulation}, i.e., to leverage the visual representation of an object to execute an operation (direct manipulation) or, if that is not possible, to at least trigger the operation based on that object. For example, connections between a node and edge class are specified through dragging an edge connector to the node, while attribute-based operations are initiated through interacting with the attribute header. We argue that this immediacy is a distinct advantage of an interactive visual wrangling solution over a programming-based approach. 

%in lieu of programming to provide users with the ability to model and reshape networks. We use the term in-place manipulation to refer to interactions that are as close to a visual representation of data as possible, while avoiding the stricter term \textit{direct manipulation}.

The interface contains three main views, shown in Figure~\ref{fig:teaser}: a network model view, an attribute view, and a network sample view, in addition to various operation-specific views.  The network model view is one of several ways users can modify the state of the network by generating new node and edge classes, and connect existing classes. The attribute view shows a table for each node and edge class, where attributes are columns and rows are instances. Table headers enable users to filter, sort, control instance labeling, and generate new node classes based on attributes. The network sample view displays a sample of the network with a concrete instance of each class in the network model. Views can be flexibly placed and scaled.

Origraph aims to make interactions for operations as close to their visual representation as possible. For example, operations such as \connect~\textbf{connecting} node classes are supported through direct manipulation of the network model; similarly, operations that utilize attributes are accessible from the attribute headers.

%Origraph uses a layout manager to allow users to manage views through resizing, drag-and-drop layout changes, and even popping the views out to entirely new windows.

\subsection{Network Model View}

\begin{figure}[b]
  \centering
  \vspace{-5mm}
    \includegraphics[width=0.4\linewidth]{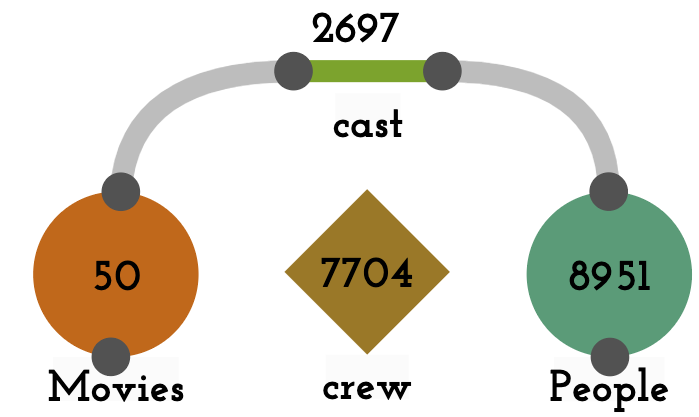}
   \vspace{-3mm}
  \caption{A simple network model for the movies dataset. Node classes (movies, people) are represented as circles, edge classes as lines (cast), and generic classes as diamonds (crew). The cast edge class connects movies and people.}   \label{fig:model_view}
\end{figure}

The network model view serves to represent all classes and their relationships, and to initiate modeling operations. The network model is represented as a node-link diagram, as shown in Figure~\ref{fig:model_view}. Node classes are represented as circles and edge classes as lines. Generic classes (before they are assigned to be nodes or edges) are represented as diamonds. We chose these encodings because they are consistent with the mental model most analysts have about nodes and edges. Recall that edges can have arbitrarily many and complex attributes, just as nodes, and that they can exist independent of nodes, which implies that we need to interact with edges in the same ways as with nodes. Our final edge encoding enforces that a segment of the edge is always horizontal and allows a connection to a node on each side (notice that a segment of the cast edge is horizontal in Figure~\ref{fig:model_view}). The horizontal segment ensures readable labels (no tilted text) and a simplified way of interacting with edges (no rotation). A downside of these restrictions is that effective layouts can be challenging with off-the-shelf algorithms, which is why we rely mostly on manual layouts. We also experimented with other glyphs for edges that are more flexible in terms of where we can draw connection lines. We, however, found it more important to use the edge metaphor than to enable better automatic layouts.

The number of items in each class is displayed with labels. We also experimented with graphical encodings for the number of items, but found that significantly different class sizes are common, resulting in unbalanced visuals and difficulties in interacting with the classes. Each class is assigned a unique color and a class label. Both are used consistently across all views for elements associated with that class. Although qualitative color scales are limited in the number of distinguishable colors, we observed that most cases do not exceed six or seven classes. For more than eight classes, we use gray as the color and fall back to labels and interactive highlighting. 

Nodes and edges have handles that can be used to initiate \connect~\textbf{connection} and \edgeProjection~\textbf{edge projection} operations by dragging the handle of one class to the other. \convert~\textbf{Conversion} and \toggleDirection~\textbf{direction changes} are available in place in the network model view.

\subsection{Attribute View}

The \textbf{attribute view} uses a tabular layout with multiple tabs to represent classes, attributes, and items. Each class is displayed in a separate table/tab, with a column for each class attribute and a row for each item. In contrast to other tools~\cite{liu_ploceus:_2014, heer_orion:_2014}, we chose to show not only the attributes and data types, but also the whole table, because we believe that it is important to be able to quickly assess the full dataset when making wrangling decisions. Our attribute view shows the data in cells as numbers or strings, but we consider visual encodings for future versions~\cite{furmanova_taggle:_2017}. The attribute table can also handle nested data structures, such as arrays and objects, which we encode using curly or square brackets for objects and arrays, respectively, enclosing a number indicating the size of the data structure. 

Column headers serve as the interface for initiating operations that are based on attributes, including the modeling operations \promote~\textbf{promotion} and \facet~\textbf{faceting}, the item operations \filter~\textbf{direct} and \connectiveFilter~\textbf{connectivity-based} filtering, as well as the attribute operations for \deriveAttribute~\textbf{in-class} and \deriveConnectedAttribute~\textbf{across-class} attribute derivation. 

% The attribute view also serves for seeding the network sample view with nodes or edges of interest and to control the appearance of the network in the sample view. The relationships between attributes and the network topology can be explored using linked highlighting between these views.  

%Although using a table to facilitate attribute-based operations requires horizontal scrolling, it is an appropriate fit because it includes the usability benefits of a familiar interface, together with a small number of visible sample rows that provide context for the kinds of values that each attribute contains.

\subsection{Network Sample View}

The network sample view renders a force-directed node-link diagram containing a sample of the network in its current state. The nodes are colored according to their class; node labels are shown on hover, and are rendered persistently for selected nodes and their neighbors. The attribute view and network sample view are linked through highlighting. Which attribute to use as a label can be changed in the attribute view. 

By default, we sample from node and edge classes while ensuring that the sampled items are connected. We achieve this by using the depth-first sampling approach described by Hu and Lau~\cite{hu_survey_2013}, but slightly modify it to balance sampling across classes.
Alternatively, users can seed nodes or edges of interest using controls in the network sample view, or add specific items from an attribute table. To see the relationships of a specific node, its neighbors can be added on demand. 

Our strategy of working with network samples and selected items is based on the assumption that most networks wrangled with Origraph exceed what can sensibly be drawn using a simple node link layout. Sampling is appropriate when the goal of the analyst is to quickly judge the effect of wrangling operations, whereas the approach of selectively querying and expanding a subgraph allows analysts to address focused network tasks, where the details of specific node and edge neighborhoods and their attributes matter~\cite{nobre_juniper:_2019}.   

%Although Origraph is not intended for exhaustive network visualization or analysis on its own, it is still important to give users a sense for whether their high-level modeling operations are successful at an instance level. 

\subsection{Operation Views}

In addition to the views that represent the state of the network and the underlying data, Origraph provides several views designed to aid analysts in executing operations. 

% TODO: probably get rid of subsubsections...?
\subsubsection*{Connection Support Interface}

The \connect~\textbf{connect} operation matches items in two classes. Origraph supports node-edge or node-node connections. Node-edge connections connect the nodes to only one ``side'' of an edge and are commonly followed by a similar operation on the other side. For example, when treating movie roles as edges, we connect that edge first to actors and then to movies using two connect operations. Node-node operations implicitly introduce a new edge class. 

\begin{figure}[tb]
  \centering
    \includegraphics[width=\linewidth]{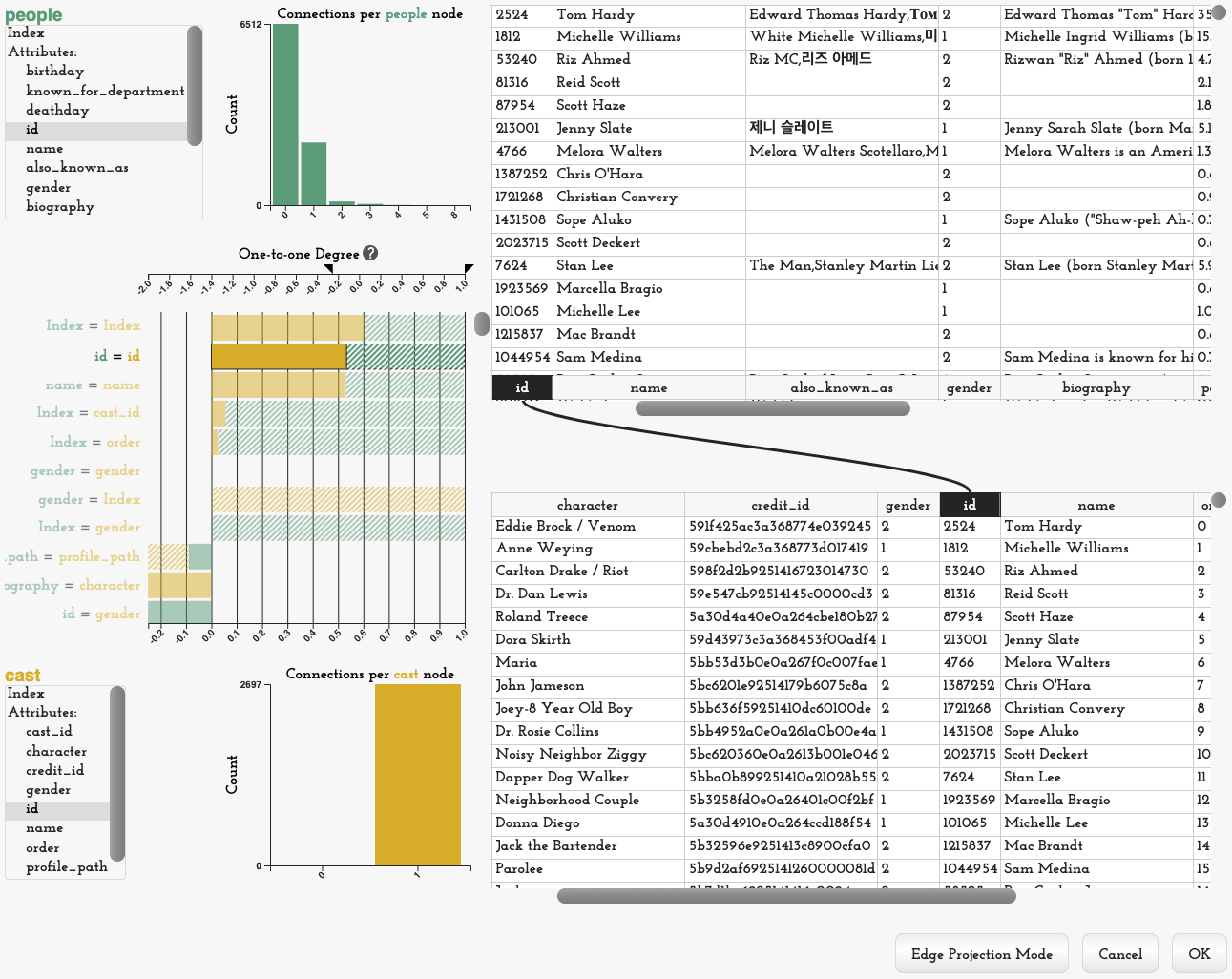}
    \vspace{-5mm}
  \caption{The connection support interface aids in finding the right connections between attributes of two different classes. The raw data of each class is shown on the right. The left side shows a score for different attribute combinations. Here the column named `id' is selected in both classes (people and cast). The histograms at the top and the bottom show the degree distribution for the selected attribute.}
  \vspace{-5mm}
    \label{fig:connect_interface}
\end{figure}

Connecting two classes can be difficult, especially if the attribute tables are large, and inconsistent column labels are used. Origraph supports this process through a visual interface, shown in Figure~\ref{fig:connect_interface}. The interface calculates all pairwise matches between the source and target classes for all attribute combinations and the index of the items. Based on that information, we calculate a heuristic score for every attribute combination between the items of two classes ($src$ and $trg$) as follows:

\[
score_{n,m} =
\mathlarger{\sum}_{class \in \{src, trg\}}
\frac{
    \sum\limits_{item \in class}
    \begin{cases}
        \frac{1}{deg_{n,m}(item)}, & \text{if }deg_{n,m}(item) > 0 \\
        -1,  & \text{otherwise}
    \end{cases}
}{|class|}
\]

Here, $n$ is the attribute in the $src$ class, $m$ in the $trg$ class, and $deg_{n,m}(item)$ is the degree of the item when connecting between the classes based on these attributes. The heuristic is designed to give a greater weight to 1-1 matches and slightly discounts matches to nodes with higher degrees. Discounting nodes with higher degrees avoids issues with cases where a small set of categorical labels that are identical between the two classes produces many matches. The heuristic penalizes missing connections and normalizes the final scores to the class size. The heuristic results in a score of 2 if each item in the source class matches exactly one item in the target class and \newText{there are no mismatches}, and -2 if no matches are found.

The contribution of each class to this heuristic is shown in the interface as a stacked bar chart. Solid bars encode the final score, and shaded bars encode one-sided scores where the overall score is smaller than the score of one class. Additionally, histograms show the degree distribution of the items. Figure~\ref{fig:connect_interface} shows connections for people to a list of cast members. The people class contains many individuals who are crew but not cast members, but each cast member has a match in the people table. The people degree distribution shows that most people acted in no movies (the crew), followed by many in a single movie, and much fewer in multiple movies. The heuristic score shows that the highest score matches the indices of the two classes, which is not correct. The score between the two id attributes is slightly lower. The relationship between these attributes is also shown as a link between the tables.

\subsubsection*{Path-Based Operations}

Several operations, including \edgeProjection~\textbf{edge projection}, \deriveConnectedAttribute~\textbf{deriving attributes based on connectivity}, and \connectiveFilter~\textbf{connectivity-based filtering}, require analysts to specify conditional paths in a network. For example, to connect an actor with a direct edge to a production company, as illustrated in Figure~\ref{fig:op_project_edges}, analysts have to specify the path from actor to role to movie to production company. 

\begin{figure}[tb]
  \centering
    \includegraphics[width=\linewidth]{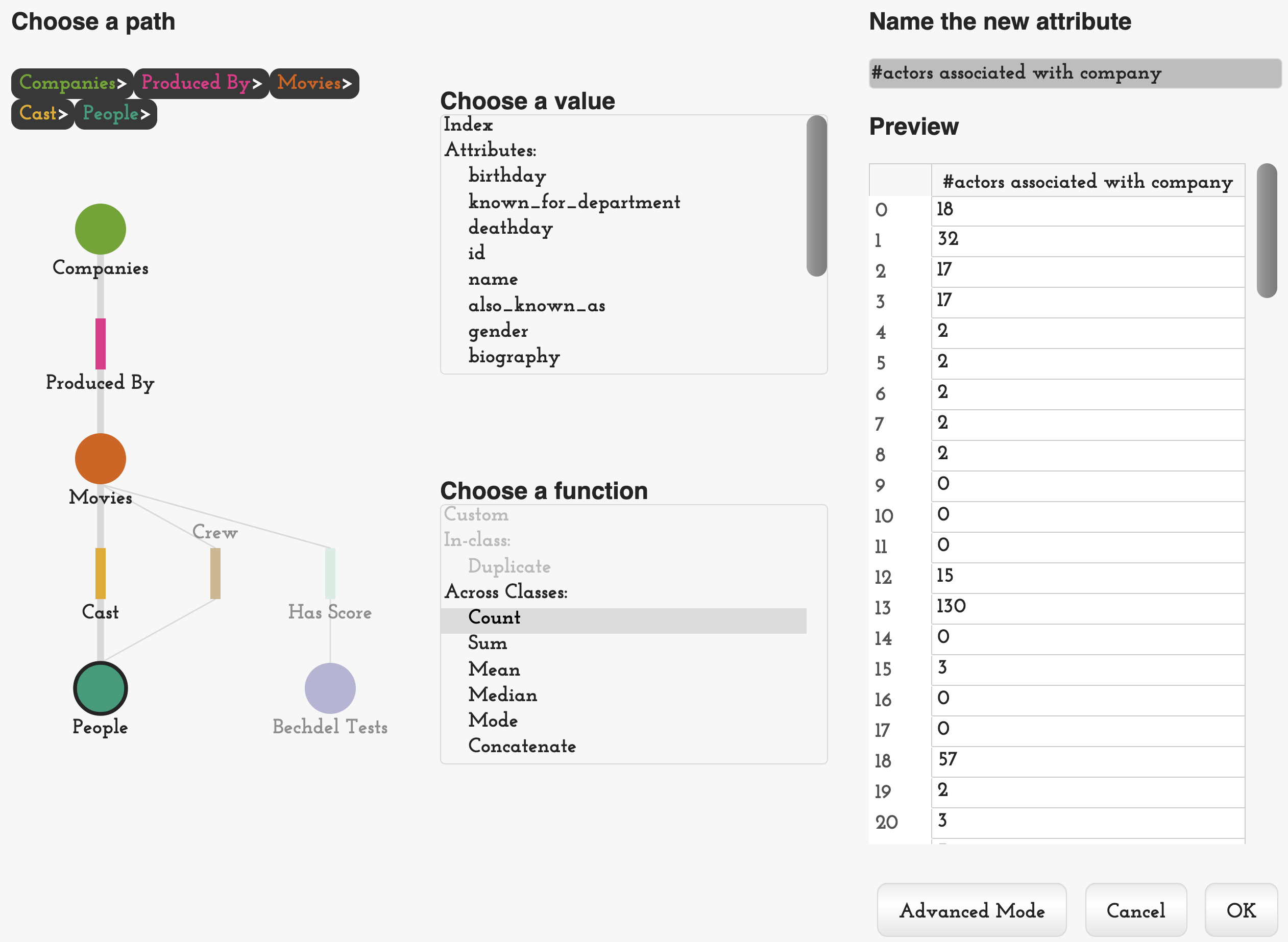}
      \vspace{-5mm}
  \caption{Path selection (left) and apply function (right) interface showing a \textbf{derive connected attribute} operation counting the number of actors associated with a production company. A preview of the results is shown on the right.}
  \vspace{-5mm}
    \label{fig:project_interface}
\end{figure}

To simplify selecting these paths, Origraph includes an interface that displays all paths across node and edge classes starting from a selected node, as shown in Figure~\ref{fig:project_interface}. Analysts can select specific paths (including loops), which are shown in a breadcrumb interface at the top. These paths are then used as input to the aforementioned operations.

%it possible for analysts  to take advantage of connected information, Origraph includes a node-link view of the network model, with a breadcrumb, that lets users specify a path across the model's topology, shown in Figure~\ref{fig:movies_derive_template_attr}. User-defined paths can wander anywhere in the graph, and may even contain repeated cycles; for example, a Person-Cast-Movie-Cast-Actor-Cast-Movie-Cast-Actor path, coupled with an equality filter of ``Kevin Bacon'' on the final Actors' name attribute, would filter the Actors to the set that has a Bacon Number of 2.

\subsubsection*{Specifying Functions}

The two operations concerned with attribute derivation (\deriveAttribute~\textbf{in-class} and \deriveConnectedAttribute~\textbf{connectivity-based} attribute derivation); and the filter operations (\filter~\textbf{direct} and \connectiveFilter~\textbf{connectivity-based} filtering) are based on evaluating functions over one or multiple values. Here, the tension is most acute between supporting operations that do not strictly require programming, yet could benefit significantly from its expressiveness. To address this balance, each of these operations has two modes. One is a standard, non-programming mode that provides sets of common functions such as count, sum, or median for attribute derivation; or simple less-than, equality, or greater-than comparisons for filtering (see Figure~\ref{fig:project_interface}).

An advanced mode drops analysts into a pre-populated code template that implements the standard function they had previously chosen. They can then adapt this function as needed. For example, Figure~\ref{fig:movies_derive_attributes} shows the computation of a gender bias ratio, which was minimally adapted from auto-generated code for computing the median. %In contrast to query languages such as SQL, users only need to write code where more expressiveness is needed; the rest can be performed interactively.

%\begin{figure*}[htbp]
%  \centering
%    \includegraphics[width=\linewidth]{figures/implementation/4.pdf}
%  \caption{\todo{change this figure to a more complete model} Origraph's internal architecture uses two layers of abstraction to map between raw data and user-driven models. Users interact directly with the interpreted network model directly in Origraph's interface, while Origraph maintains an internal table network that tracks where classes and attributes exist in the raw data, so that values can be derived as needed.}
%    \label{fig:architecture}
%\end{figure*}

\section{Implementation}
Origraph is a client-only web application, published under a permissive open-source license. The source code is available at \url{https://github.com/origraph/origraph.github.io}. %The visual interface is  implemented with D3~\cite{bostock_d3:_2011}.  
%^A SENTENCE ABOUT libraries etc (typescript and what not)

Origraph is designed to help analysts create a set of rules for reshaping a graph in a lightweight, flexible interface. In its current form, it scales to medium-sized networks (up to tens of thousands of nodes) that can fit in memory. However, Origraph's interface is built on top of an independent graph processing library \texttt{origraph.js}, available at \url{https://github.com/origraph/origraph.js}. Ultimately, our goal is to support interactive operations with an in-memory sample of a large graph, and then export a script capable of applying users' rules to much larger datasets, similar to the strategy pioneered by existing tabular data wrangling tools~\cite{kandel_wrangler:_2011,verborgh_using_2013,trifacta_trifacta_2012}.

The underlying library interacts with raw data using two different layers of abstraction: a \textbf{table network} layer, underneath an \textbf{interpreted network model} layer. Similar to the schema of a relational database, the table network is a lazily evaluated specification of raw and derived tables, and any connections between them. The interpreted network layer maps tables in the table network to classes in the network model: both node and edge classes must map to a specific target table. This way, edges in the graph are always guaranteed to support features, such as edge attributes, that existing network modeling tools~\cite{liu_ploceus:_2014,heer_orion:_2014} cannot represent. Furthermore, edge classes reference up to two paths through the table network, from the target edge table to its connected node tables. Distinct paths between tables imply that edge classes can be fully or partially disconnected, and these states are reflected in Origraph's interface. Keeping these two layers separate allows for free-form reinterpretation of the underlying tables as node or edge classes.

\section{Input and Output}

As for any data wrangling tool, input and output are important considerations. Our design goal for input is to ingest data formats from ``in-the-wild'' data sources and to avoid the need for pre-processing. To this end, Origraph supports tabular data that can contain explicit node and link lists; alternatively links can be inferred based on attributes.

Another input data type is hierarchical, specifically JSON, which is a common format for API responses from numerous online services. The hierarchy commonly contains multiple levels of items that can be individually represented as nodes and edges. A movie object, for example, can contain an array of all cast members, themselves represented as complex objects. Origraph implements special \textbf{unroll} and \textbf{expand} operations to convert these nested structures into separate classes.

Origraph currently exports d3.js-style JSON, zipped CSV files, and GEXF for analyzing the wrangled graphs in off-the-shelf graph visualization tools such as Gephi or Cytoscape. The prototype comes with multiple datasets that can be loaded in raw or pre-assembled format. 

\section{Use Cases}
\label{sec:use-cases}

Here we demonstrate how Origraph can be used to wrangle two network datasets. The data for both cases was retrieved from several different APIs, and no data wrangling was performed outside of Origraph. The scripts used to access the API endpoints and the resulting data can be found at \url{https://github.com/origraph/data}; each dataset is also provided as a sample in Origraph. %We highlight this to underscore Origraph's ability to handle data beyond the traditional tabular csv format, including the complex nested objects often returned from APIs, in JSON format. 
% We wrangle the original data with the goal of creating network datasets that can then be imported into external network and/or tabular visualization tools, such as Gephi, for analysis. Larger versions of each figure are available in the supplemental material.
Figures showing the full state of the interface at each step (M1--M18; Y1--Y15) are available in the supplemental material.

\subsection{Gender Bias in Movies}
\label{sec:us-movies}

The network used in this example is a network of movies, actors, crew members, and production companies. The data was retrieved from ``The Movie Database''~\cite{themoviedatabasecommunity_movie_2018}, a community-built movie and TV database.  %Each item in the dataset is associated with rich attributes. 
%For example, movies contain not only basic information such as title and year of release, but also budget, genre, spoken languages, popularity, run time, and so on. 
%Actors have attributes such as gender, popularity, place of birth, and day of death. The data was retrieved through the TMDB API. 
The analysis question we want to investigate concerns issues related to gender bias in the movie industry.

We select the 50 most popular movies (according to TMDB's internal ranking of popularity), retrieve data related to these movies, and store the data in three JSON files. The movies file contains information on the 50 most popular movies, including attributes such as movie id, title and year of release, budget, genre, spoken languages, popularity, run time, revenue, and nested objects describing the production companies involved. The people file contains information on all people in these movies and contains a unique id for each person in addition to attributes such as gender, popularity, place of birth, and day of death. The credits file contains two nested objects, one for cast and one for crew. These objects contain attributes such as roles, departments, and the ids for movies and people. In total, the datasets contains 2689 cast items, 7704 crew items, 116 production companies, 181 relationships between production companies and movies, and 8951 people.

We also retrieved a dataset containing Bechdel ratings for 7871 movies~\cite{bechdeltest_bechdel_2018}, including the 50 that we retrieved from TMDB. The Bechdel test assigns a rating from 0-3 to each movie based on three criteria: (1) the movie must have at least two women in it, who (2) talk to each other, about (3) something besides a man. A movie that fails all three criteria is assigned a score of 0; a movie that passes the test is assigned a score of 3. Supplementing the TMDB data with the Bechdel rating allows us to answer some interesting gender-related questions about the movies. 

We start the process of modeling a network by importing the raw data described above.
%\oldText{Each of the imported data files is immediately available as separate, generic classes, represented by a table in the attribute view and a diamond in the network model view (Supplementary Figure~M1). The first step in modeling the network is choosing which entities to interpret as nodes and which as edges. We start by \convert~\textbf{converting} generic movies to nodes.}
%\oldText{The raw data, however, has movie information from two different sources, but we need to combine the Bechdel movies and the TMDB movies into a single node class. To do this, we also \convert~\textbf{convert} the Bechdel movies to nodes (Figure~M2). Each class has an IMDB identifier as an attribute, however, in the TMDB dataset, those IDs have a ``tt'' prefix before each number. In order to \connect~\textbf{connect} the two node classes, we first need to \deriveAttribute~\textbf{derive an attribute} for the Bechdel movies, appending a ``tt'' string to its IMDB identifier (Figure~M3). Now, we can \connect~\textbf{connect} each movie class (Figure~M4).}
Our first objective is to combine the Bechdel and TMDB data about movies. To do this, we employ \convert~\textbf{convert}, \deriveAttribute~\textbf{derive attribute}, \connect~\textbf{connect}, and \deriveConnectedAttribute~\textbf{derive connected attribute} operations to arrive at a combined movies class with attributes from both sources (see Figures~M1--M6).
%\oldText{ for the TMDB movies, duplicating the original Bechdel score---in this case, we select the ``rating'' attribute from the Bechdel movie class, and compute the median rating (Figure~M5). Note that the general case for deriving connected attributes is a many-to-one relationship. Because the data here happens to have a one-to-one relationship, the median computation trivially copies the only connected value. Once we have copied the value, the Bechdel movie class and its connection to the TMDB movies are no longer needed, and both classes can be deleted (Figure~M6).}
%\begin{figure}[b]
%  \centering
%    \includegraphics[width=\linewidth]{figures/movies_derive_template_attr.png}
%  \caption{Deriving an attribute across connected classes}
%    \label{fig:movies_derive_template_attr}
%\end{figure}
We then unroll nested cast and crew objects as distinct classes (Figure~M7), \convert~\textbf{convert} people to nodes, cast and crew to edges (Figure~M8), and then \connect~\textbf{connect} both cast and crew to people and movies (Figure~M9). \newText{At this stage, we have modeled a basic network; movies and people are connected through cast and crew edges, and each movie has an associated Bechdel score. This dataset might suffice to answer basic questions concerning which people participated in movies with high or low Bechdel scores; however, an analyst might also be interested in gender bias at the level of production companies.}

\newText{To treat production companies as distinct entities,} we scroll through the movie attributes to ``production\_companies''. The values in this column are arrays of objects, so we unroll them into their own class (Figure~M10). The unroll operation also connects the new items to the movie nodes they originated from. We rename the newly created class ``produced by'' to better reflect the meaning of this class in our model.

\oldText{At this stage, because the same production company is associated with multiple movies, we notice the presence of duplicate rows, showing information for the same company (Figure~M11).}
%\begin{figure}[htbp]
%  \centering
%    \includegraphics[width=\linewidth]{figures/movies_network_modeling.png}
%  \caption{State of the network after importing the data, converting movies to nodes, unrolling cast and crew into their classes and using them to connect people to movies, and unrolling production companies into their own class.}
%    \label{fig:movies_network_modeling}
%\end{figure}
We leverage the \promote~\textbf{promote} operation on the company ``name'' attribute to create a new class with unique company names. We then rename the newly created class ``Companies'' (Figure\newText{s~M11--}M12). However, companies are now connected to movies only indirectly via \oldText{the }``produced by'' nodes. Semantically, it would be more meaningful to connect companies directly to movies via an edge, which we can achieve with the \convert~\textbf{convert} operation, applied to the ``produced by'' node class (Figure~M13). Since Origraph supports rich edge attributes, these edges preserve all the attributes of the original data.

%\begin{figure}[htbp]
%  \centering
%    \includegraphics[width=\linewidth]{figures/movies_production_companies.png}
%  \caption{State of the network after promoting the companies attribute to remove duplicates.}
%    \label{fig:movies_production_companies}
%\end{figure}

% Next, we import the Bechdel scores, make them nodes, and link movies to their Bechdel scores based on the movie ID. However, the ID in the movie table is formatted slightly differently in these two independent datasets. 
% We use the \deriveAttribute~\textbf{derive a new attribute} operation in the Bechdel table to re-format the id to match with the IDs used in the movie nodes. We need a custom function to achieve this, which we specify with the provided code editor (shown in Figure \ref{fig:movies_derive_attributes}). Once the id columns match, we can \connect~\textbf{connect} movies to their Bechdel scores, resulting in the final network model seen in Figure \ref{fig:teaser}.

\begin{figure}[t]
  \centering
 
    \includegraphics[width=\linewidth]{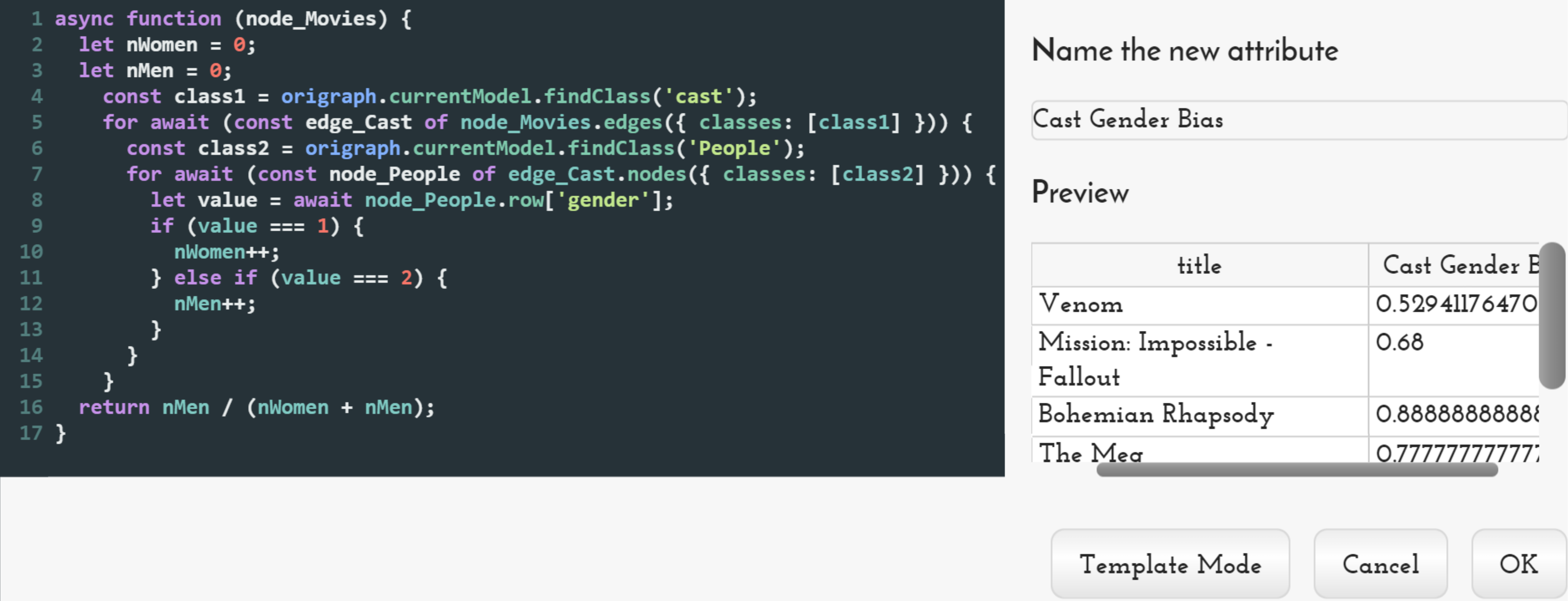}
          \vspace{-5mm}
  \caption{Interface for deriving a custom attribute using code. In this case, the user adapts an auto-generated template for computing the mean value, changing eight lines of code: the counter initialization lines (2,3); the \textit{if} conditionals and increments (9-13); and the return statement (16).}
         \vspace{-5mm}
    \label{fig:movies_derive_attributes}
\end{figure}

With \newText{this iteration of} the \oldText{initial} network model built, we are ready to reshape the network to \oldText{answer the}\newText{support} analysis questions regarding gender bias for each movie. To do this, we compute a new attribute in the movie table. Similar to the earlier task of copying Bechdel scores, this operation requires information from connected classes, so we use the \deriveConnectedAttribute~\textbf{connectivity-based attribute derivation} operation. In order to calculate the gender bias for actors in a movie, we must access the gender of all the people who are connected to the movie via a cast edge. In this example, we are interested in the ratio of men to women, so we select gender from the list of attributes, choose the mean computation that approximates what we want to compute (Figure~M14), and then select the advanced mode, which allows us to modify the mean code to compute the ratio of men to women (Figures~\ref{fig:movies_derive_attributes} and M15). A preview table on the right shows sample values to ensure we are computing the attribute correctly.

Once the gender bias has been computed, we sort on this attribute in the table, which reveals that of the movies in this dataset, ``Ocean's Eight'' is the one with the highest ratio of women to men (Figure~M16). Interestingly, the movie with the highest gender bias, ``BlacKKKlansman'', passes the Bechdel test, with a score of three, which suggests that one metric (or both) possibly oversimplifies the concept of gender bias. To explore the relationship between each metric, we can export the movies class as a CSV file, and visualize Cast Gender Bias, Crew Gender Bias, and Bechdel scores in a scatterplot matrix (Figure~M17). %notebook~\cite{bostock_observable_2018} .

%\begin{figure}[t]
%  \centering
%    \includegraphics[width=.8\linewidth]{figures/movies_splom.png}
%  \caption{Scatterplot matrix showing the relationships (or lack of relationships) between Bechdel ratings (a score of 3, shown in \greentext{green} means a movie passes the test) and our cast and crew gender bias metrics (high values indicate a predominantly male cast/crew).}
% \label{fig:movies_splom}
%\end{figure}

A follow-up question is to find out which actors tend to be cast in movies with a lower gender bias. We can explore the local connections around movies of interest using Origraph's sample interface (Figure~\ref{fig:teaser}), or export finished network dataset to a more dedicated network visualization tool such as Gephi (Figure~M18).

%\begin{figure}[htbp]
%  \centering
%    \includegraphics[width=\linewidth]{figures/movies_gephi.png}
%  \caption{Exploring the full movie network in Gephi.}
% \label{fig:movies_gephi}
%\end{figure}

\subsection{Money and Political Support for the War in Yemen}

Our second use case focuses on a network of current US senators, voting behavior on the recent bill regarding US support of the Yemen war, donors of these senators, and their social media statements related to this issue. This topic has received considerable attention in the media~\cite{maza_republican_2018, freeman_meet_2018} and serves as an interesting use case to demonstrate the ability of Origraph to connect data from disparate sources to tell a compelling story. The bill in question (session 2, roll call 250) was to determine whether the US should remove military support from the war in Yemen. The final vote count was in support of removing support. Yet it is still interesting to model a network that can address questions related to the votes of senators and their connections to donors. One such question is whether senators who voted against the Yemen bill were financially supported by a specific subset of donors with an interest in the conflict. A related question is whether senators were more or less vocal in their support or opposition, which we can estimate based on press releases and tweets.

We obtained the data from the ProPublica Congress API~\cite{propublica_propublica_2018} and through the Twitter API~\cite{twitterinc._twitter_2018}. Again, we did not perform any pre-processing on the datasets retrieved from these APIs. 
The raw data files imported in this example include: information for all current senators including gender and political party, the way each senator voted for the bill on Yemen, all press releases made by members of the senate in relation to the Yemen bill, all FEC reports about donations made in 2018 that either support or oppose a member of the senate, and all tweets made by senators from Nov.\ 28 to Dec.\ 4,\ 2018.

Once the raw files have been loaded (Supplementary Figure~Y1), we \convert~\textbf{convert} senators, votes, campaign contributions, press releases, and tweets to nodes (Figure~Y2). \oldText{In order to \connect~\textbf{connect} senators to their tweets, we must first extract the twitter user information from each tweet. User information is stored as nested objects within each tweet; we generate a new class with all users by expanding the user attribute in the tweet class (Figure~Y3), which automatically connects the instances of the newly created user class to the original tweets (Figure~Y4).} We \oldText{can now}\connect~\textbf{connect} senators to their twitter accounts \oldText{based on their screen name, an attribute of the senators class}\newText{by expanding nested information} (Figure\newText{s~Y3--}Y5). Because we ultimately care about connecting senators to their tweets, we can abstract away the twitter account class by \convert~\textbf{converting} it to an edge. We now have a model of senator nodes, connected directly to their tweets (Figure~Y6). The next step involves \connect~\textbf{connecting} senators to their votes, their press releases, and any donations made toward or against their campaign. This step can be done directly with the connect interface, as press releases, votes, and donations all have an attribute that directly references the senator ids (Figure~Y7). We are particularly interested in distinguishing between senators who voted for or against the bill, so we \facet~\textbf{facet} the votes node into separate ``Yes'' and ``No'' classes before connecting them (Figure~Y8).

The last step in modeling this network involves extracting the individual donation committees. Since several donations are made by the same committee, we first want to extract all unique committees from the donations class. We leverage the \promote~\textbf{promote} operation on the ``committee name'' attribute, which generates a new ``Donor Committees'' class, with edges connecting each committee to their donation in the ``Donations'' class (Figure~Y9).

Because we are interested in which committees support or oppose certain senators, we \convert~\textbf{convert} the donations node into an edge (Figure~Y10), and then \facet~\textbf{facet} the edge into contributions that support and those that oppose a senator (Figure~Y11). 

\begin{figure}[t]
  \centering
    \includegraphics[width=\linewidth]{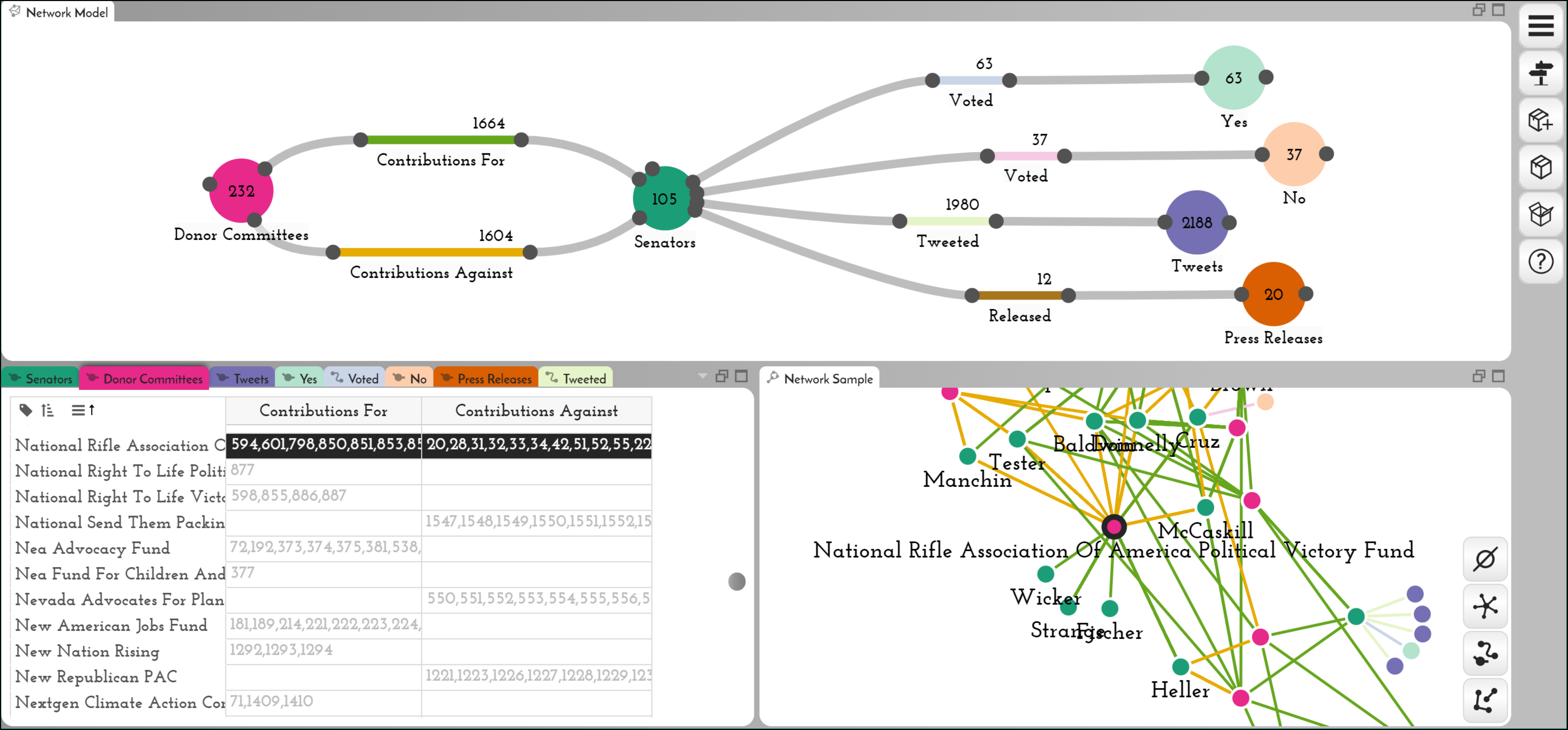}
      \vspace{-5mm}
    \caption{A political donor, vote, tweet, and press release network. Senators are connected to committees that donated to either support or oppose them. They also have edges connecting them to how they voted on the Yemen bill, their tweets, and press releases.}
      \vspace{-5mm}
    \label{fig:yemen}
\end{figure}

With this network modeled, as shown in Figure~\ref{fig:yemen}, we can turn to the initial questions regarding relationships between donors and vote outcomes on the Yemen bill. Our current network connects donor committees to Yes and No votes through specific senators. An analyst may be less interested in specific senators than relationships between funding sources and votes; consequently, the intermediate senators are a hindrance in this state of the network. To remedy this, we can \edgeProjection~\textbf{project new edges} that connect donor committees directly to vote classes (Figure~Y12), resulting in a new edge class that is a reasonable proxy for donor interest in the conflict. Exporting the new projected edges, with donor and vote nodes, into Gephi allows us to explore the donor interest network (Figure~Y13). We discover that, even though many republicans voted Yes on this bill, the NRA supported only Republican senators who voted No, and opposed senators who voted Yes. Although this pattern in connectivity does not directly prove NRA support for the Yemen conflict, it could suggest a point that an analyst or journalist might want to investigate further.

Having incorporated Twitter and press release data, we may also be interested in whether specific senators were vocal about the conflict. We can derive attributes on senators to ascertain whether their tweets contain relevant ``Khashoggi,'' ``Saudi,'' or ``Yemen'' strings, as well as whether or not a senator issued a press release about their vote (Figure~Y14). These new attributes can be interpreted as set relationships---yet another data model---that might be better explored in a tool such as UpSet~\cite{lex_upset:_2014}. Exporting the derived data to UpSet reveals that, with the exception of only three senators, senators who voted No were relatively silent about their votes (Figure~Y15).

%\begin{figure}[htbp]
%  \centering
%    \includegraphics[width=0.8\linewidth]{figures/politics_project_edges.png}
%    \caption{Projected edges bypass senators, and connect donors directly to votes, including each permutation of whether they supported or opposed senators who voted Yes or No. Classes are also assigned similar colors, based on similar semantic meaning, while color is removed from unused classes.}
%    \label{fig:politics_project_edges}
%\end{figure}

%\begin{figure}[htbp]
%  \centering
%    \includegraphics[width=\linewidth]{figures/politics_gephi.png}
%    \caption{Projected edges, donors, and votes are exported to Gephi. We highlight the NRA, as it did not donate to any senators who voted Yes---only to those who voted No, and donated in opposition to other senators who voted Yes.}
%    \label{fig:politics_gephi}
%\end{figure}

% \begin{figure}[t]
%   \centering
%     \includegraphics[width=1\linewidth]{figures/politics_final_network.png}
%   \caption{Network model for politics and twitter data}
%     \label{fig:politics_final_network}
% \end{figure}

\section{Discussion}
\label{sec:discussion}

Data abstractions often change in response to evolving exploration and analysis tasks. The shape of the data is a proxy for an analyst's real-world problem, and must be continuously assessed and refined~\cite{fisher_making_2018}. We argue that transforming a dataset into the form best suited to answer analysis questions is essential, yet the tools currently available for wrangling multivariate network datasets are not powerful enough. The alternative of using scripting or database query languages is time consuming, requires skill sets many analysts do not have, and does not allow for rapid iterative exploration of the changes made.

To address this problem, we introduce novel operations and visual analysis interfaces to support sophisticated interactive network wrangling. By giving users the freedom to explore alternative data abstractions, tools like Origraph have the potential to become powerful thinking tools that could begin to address the open challenge of how to support visualization practitioners in exploring more diverse data abstractions~\cite{munzner_nested_2009, bigelow_reflections_2014}. Although the design space of data wrangling tools is still in its infancy, it is worth considering its breadth: identifying data wrangling as a distinct category from data analysis~\cite{kandel_research_2011} may need to be extended further to distinguish among data analysis, classical ``wrangling'', and abstraction transformation. The standard notion of data wrangling~\cite{kandel_research_2011} is often viewed through a lens of data integration, diagnosing, and cleaning data problems. In contrast, abstraction transformation is motivated by users' mental models and hence requires a distinct set of methods that support different operations. Nevertheless, each category benefits from tight integration with the others. For example, in Origraph, we integrate data analysis through visualization with wrangling, which is also imperative for checking whether wrangling operations have their intended effects.

\subsection{Comparing Origraph to a Computational Workflow}

All wrangling operations that can be executed with Origraph can also be implemented using different programming languages. However, using a programming language has multiple disadvantages: first, the skill level required to implement certain operations limits the number of individuals who can execute them significantly, making it prohibitive for data-literate non-programmers (e.g., individuals who use Excel, Tableau, or Statistics software such as SPSS) to do network wrangling. 

All modeling operations in Origraph can be executed without programming, which is noteworthy because operations that change the database schema are relatively complicated in scripting languages such as SQL. Only non-standard attribute operations require programming in Origraph: the ``apply'' functions following the split-apply-combine paradigm and advanced conditionals for filtering items. Origraph provides multiple defaults for these functions, such as sums, means, and concatenation. However, the design space of useful functions is considerable; hence Origraph allows analysts to write custom expressions. We argue, however, that \oldText{these }expressions operating on lists of attributes are much easier to write compared to, for example, operations that change the network model. Nevertheless, we plan on simplifying our scripting interface, for example, by making it more like Excel, with basic formulas, easy references to data sources, and accessible documentation.

\subsection{Comparison to Other Tools}

Origraph shares parts of the vision and functionality of Orion~\cite{heer_orion:_2014} and Ploceus~\cite{liu_ploceus:_2014}. Origraph, however, is unique and novel with regard to three aspects: (1) it supports important operations that are missing in Orion and Ploceus, (2) it treats edges as first-class objects that can be associated with complex attributes, and (3) it contains sophisticated, algorithm-supported visual interfaces that make executing complex operations easier. Orion and Ploceus also do not support nested data structures (lists and hierarchies) as attributes of the tables. 

With regards to operations, Origraph supports all operations listed in Section~\ref{sec:operation}. Ploceus and Orion do not support \convert \textbf{converting between nodes and edges}, \supernode \textbf{creating supernodes}, \connectiveFilter \textbf{connectivity-based filtering}, \toggleDirection \textbf{changing edge direction}, and \textbf{connectivity-based attribute derivation}. Although Origraph and Ploceus support \edgeProjection \textbf{projecting edges}, they do so only for immediate neighbors, whereas Origraph can project edges based on arbitrary complex paths. Notably, several of the operations that are unique to Origraph are about leveraging network structures, e.g., aggregating attributes at a certain distance from source nodes.  Supplementary Table 1 contains a comparison of these operations.
Finally, in contrast to Orion and Ploecus, the code of Origraph is open source, and the tool is available for anyone to upload datasets. 

\subsection{Evaluation}
\newText{Criteria for evaluating data wrangling systems are not well established. Similar to visualization authoring toolkits, the primary goal of a data wrangling system is to \textit{create}, rather than \textit{analyze}---consequently, a wide variety of concerns merit evaluation, such as expressiveness, creativity support, flexibility, guidance, efficiency, usability, learnability, and integration~\cite{ren_reflecting_2018}. In this work, we have focused primarily on demonstrating expressiveness through use cases, and to some extent, the system's design demonstrates creativity support, flexibility, and integration.
%We note that, as the design space for possible data abstractions may be smaller than the design space for possible visualizations, stronger, crisper expressiveness claims may be possible by simply enumerating supported abstractions and operations.
We consciously chose not to run a quantitative study comparing Origraph with other systems or methods. A comparison to existing systems, such as Ploceus and Orion, is not possible, since these tools are not available to the public. A comparison to a programming-based approach hinges on the scripting skills of the subjects; yet our target users are those who have no or minimal programming skills.
Future work is needed to evaluate concerns that we have not considered in depth, such as usability. We also believe that a discussion around how to best evaluate data wrangling systems is needed.}

\subsection{Limitations}

\paragraph{Visualization.} The current focus of Origraph is on making sophisticated data wrangling operations as easy as possible to execute. Origraph supports analysis through visualizations, but visualizations of the network and the attributes can be improved in many ways. We are planning to integrate Origraph with a general purpose multivariate network visualization system that is based on existing tools developed at our lab~\cite{nobre_juniper:_2019, kerzner_graffinity:_2017, partl_pathfinder:_2016}.  

\oldText{Cleanup and Missing Data.} \oldText{Origraph does not support operations on tabular data that analysts have come to expect from a wrangling tool.}
\newText{\paragraph{Focus on Abstraction Design.} Origraph targets the design of data abstractions, rather than the discovery, capture, curation, or creation of data~\cite{muller_how_2019}.} For example, the support for cleanup and dealing with missing data is limited. We believe that such operations are better addressed in separate tools before importing data into Origraph.

\paragraph{Scalability.} Although Origraph scales to thousands of nodes and edges, it does not scale to arbitrarily large networks. We plan on improving scalability using an approach common to visual data wrangling tools: loading a sampled dataset. Operations can then applied and refined on the sample using visual inspection. After the transformation is completed, a script is generated that can be used to process a larger dataset offline, potentially leveraging cloud infrastructure.

\paragraph{Connectivity Heuristic.} In our experiments with various datasets, we found that our heuristic always gives high scores to the right combination of attributes, but also observed that index-to-index scores are commonly highly ranked. If, for example, two datasets have the same number of rows but have nothing else in common, our heuristic would give a perfect score to the index-to-index connection. Since some datasets rely on the index of items to establish connections between datasets, we cannot generally exclude it from our computation. However, we are considering treating connectivity combinations involving an index separately from attribute to attribute connections. We observed that our heuristic can be slow with large classes. It has a runtime complexity of $\mathcal{O}(n*m*i*j)$  where $n$ and $m$ are the number of unique items in the classes (which are frequently very large), and $i$ and $j$ are the number of attributes plus the index in the classes (which are typically small). Sampling could reduce the runtime complexity to $\mathcal{O}(k*m*i*j)$, where $k$ is a user-chosen value that is much smaller than $n$ and should result in a significant speed-up with a limited loss of accuracy.

\section{Conclusions and Future Work}

By implementing the set of operations that we have identified, Origraph is a first steps toward allowing data-literate non-programmers to wrangle networks to answer important questions in their area of inquiry and to produce visualizations they can use to communicate their findings. Origraph is designed to ingest raw data, e.g., in the form of JSON retrieved from an API, and then enable analysts to wrangle the data into a network that corresponds to their mental model. Build-in visualization capabilities enable analysts to quickly iterate between analysis and modeling. Once a network is in the desired form, it can be either investigated within Origraph or exported for analysis in more sophisticated multivariate network visualization tools. Through two use cases, we have demonstrated Origraph's expressive potential. %Together, these contributions have implications for data abstraction transformation tools in general.

We hope to explore and augment Origraph through ongoing deployment, testing, and redesign. These efforts will include more exhaustive evaluations with users testing and improving the tool's usability; exploring the usefulness of each operation and what operations may be useful beyond the ones that we have identified; and identifying gaps in its expressiveness. Additionally, we plan to integrate more representative sampling and improve scalability; to give users access to more sophisticated algorithms, such as seriated instance ordering, or clustering-based group assignments; to add provenance features to make it possible to edit, audit, share, document, and replicate users' transformation processes; and to integrate more closely with data cleaning and multivariate graph visualization to more fully support end-to-end analysis workflows.

\section{Acknowledgments}

This work was funded by the National Science Foundation (OAC 1835904, IIS 1350896, IIS 1751238) and by the Utah Genome Project.

\bibliographystyle{styles/abbrv-doi}

\bibliography{Origraph.bib}
\end{document}